\documentclass[11pt]{article}
\usepackage[utf8]{inputenc}
\usepackage{hyperref}
\usepackage{amsfonts}
\usepackage{amsmath}
\usepackage{amssymb}
\usepackage{indentfirst}
\usepackage{graphicx}
\usepackage{cite}
\usepackage{color}
\usepackage{xcolor}
\numberwithin{equation}{section}
\allowdisplaybreaks
\makeatletter

\begin{document}

\begin{titlepage}
\vskip 4cm

\begin{center}
\textbf{\LARGE{Three-dimensional Maxwellian Extended Newtonian gravity and flat limit}}
\par\end{center}{\LARGE \par}

\begin{center}
	\vspace{1cm}
	\textbf{Patrick Concha}$^{\ast}$,
    \textbf{Lucrezia Ravera}$^{\ddag,\dag}$,
	\textbf{Evelyn Rodríguez}$^{\star}$,
    \textbf{Gustavo Rubio}$^{\ast}$,
	\small
	\\[6mm]
	$^{\ast}$\textit{Departamento de Matemática y Física Aplicadas, }\\
	\textit{ Universidad Católica de la Santísima Concepción, }\\
\textit{ Alonso de Ribera 2850, Concepción, Chile.}
	\\[3mm]
    $^{\ddag}$\textit{DISAT, Politecnico di Torino, }\\
	\textit{Corso Duca degli Abruzzi 24, 10129 Torino-Italy.}
	\\[3mm]
	$^{\dag}$\textit{INFN, Sezione di Torino, }\\
	\textit{Via P. Giuria 1, 10125 Torino-Italy.}
	\\[3mm]
	$^{\star}$\textit{Departamento de Ciencias, Facultad de Artes Liberales,} \\
	\textit{Universidad Adolfo Ibáñez, Viña del Mar-Chile.} \\[5mm]
	\footnotesize
	\texttt{patrick.concha@ucsc.cl},
    \texttt{lucrezia.ravera@polito.it},
	\texttt{evelyn.rodriguez@edu.uai.cl},
    \texttt{gustavo.rubio@ucsc.cl},
	\par\end{center}
\vskip 20pt
\centerline{{\bf Abstract}}
\medskip
\noindent In the present work we find novel Newtonian gravity models in three spacetime dimensions. We first present a Maxwellian version of the extended Newtonian gravity, which is obtained as the non-relativistic limit of a particular $U(1)$-enlargement of an enhanced Maxwell Chern-Simons gravity. We show that the extended Newtonian gravity appears as a particular sub-case. Then, the introduction of a cosmological constant to the Maxwellian extended Newtonian theory is also explored. To this purpose, we consider the non-relativistic limit of an enlarged symmetry. An alternative method to obtain our results is presented by applying the semigroup expansion method to the enhanced Nappi-Witten algebra. The advantages of considering the Lie algebra expansion procedure is also discussed.

\end{titlepage}\newpage {} 

\tableofcontents

\noindent\rule{162mm}{0.4pt}

\section{Introduction}

Non-relativistic (NR) geometry has been of particular interest in condensed matter systems \cite{Son:2008ye,Balasubramanian:2008dm,Kachru:2008yh,Bagchi:2009my,Bagchi:2009pe,Christensen:2013lma,Christensen:2013rfa,Taylor:2015glc} and NR effective field theories \cite{Son:2013rqa,Hoyos:2011ez,Geracie:2015dea,Gromov:2015fda}. In such geometric framework, there has been a growing interest in studying NR gravity theories \cite{Duval:1983pb,Duval:1984cj,Duval:2009vt,Andringa:2010it,Banerjee:2014nja,Banerjee:2016laq,Bergshoeff:2017dqq,Aviles:2018jzw,Aviles:2019xed,Chernyavsky:2019hyp,Concha:2019lhn,Harmark:2019upf,Hansen:2020pqs,Ergen:2020yop,Concha:2020sjt} which could be relevant in approaching realistic theories.

It is well-known that standard Newtonian gravity, describing the physical gravitational force, can be geometrized using what is known as Newton-Cartan (NC) geometry. Nevertheless, an interesting open issue has been to construct an action principle for Newtonian gravity which is based on the Bargmann algebra \cite{LL,Grigore:1993fz,Bose:1994sj,Duval:2000xr,Jackiw:2000tz,Papageorgiou:2009zc,Bergshoeff:2016lwr}. The aforementioned action has recently been presented in \cite{Hansen:2018ofj}. Such construction has required, however, an underlying symmetry algebra which differs from the usual Bargmann algebra as it includes three additional generators. Subsequently, a three-dimensional Chern-Simons (CS) formulation based on this particular extension of the Bargmann algebra has been introduced in \cite{Ozdemir:2019orp}. For this formulation it was necessary to introduce a central extension in order to have a non-degenerate invariant bilinear form. Such NR gauge invariant theory has been denoted as extended Newtonian gravity and has been obtained as a contraction of a bi-metric model being the sum of the Einstein gravity in the Lorentzian and Euclidean signatures. As pointed out in \cite{Ozdemir:2019orp}, although the extended Newtonian gravity is based on the central extension of the algebra that leads to an action for Newtonian gravity, it is distinct from the Newtonian gravity action constructed in \cite{Hansen:2018ofj}. In addition, both models differ at the level of matter coupling. In fact, the matter coupling of the extended Newtonian gravity theory admits backgrounds with non-trivial curvature whenever matter is present, which makes it resemble to the case of matter-coupled extended Bargmann gravity \cite{Bergshoeff:2016lwr}.

The incorporation of a cosmological constant to the extended Newtonian gravity has then been presented in \cite{Concha:2019dqs} by introducing by hand a new NR symmetry called exotic Newtonian algebra. The respective relativistic counterparts of the extended Newtonian gravity and its exotic version have been recently studied in \cite{Bergshoeff:2020fiz}. In particular, the authors of \cite{Bergshoeff:2020fiz} showed that the extended Newtonian gravity model and the exotic one appear as  NR limits of relativistic CS gravity theories based on the co-adjoint Poincaré and co-adjoint AdS algebra, respectively.

Another extension of the Bargmann algebra is given by the recently introduced Maxwellian extended Bargmann (MEB) algebra allowing to construct a NR CS gravity action in presence of a covariantly constant electromagnetic field \cite{Aviles:2018jzw} without cosmological constant. Such NR model has been obtained as a NR limit of the [Maxwell]$\oplus \, \mathfrak{u}(1) \oplus \, \mathfrak{u}(1) \oplus \, \mathfrak{u}(1)$ CS gravity. As in the extended Bargmann case \cite{Bergshoeff:2016lwr}, the $U(1)$ enlargement of the relativistic counterpart is crucial to define a proper NR limit leading to a finite NR gravity action based on a non-degenerate bilinear form. The relativistic Maxwell algebra has been introduced to describe a Minkowski background in presence of a constant electromagnetic field \cite{Schrader:1972zd,Bacry:1970ye,Gomis:2017cmt}. Such symmetry and its generalizations have been useful to recover standard General Relativity without cosmological constant as a limit of CS and Born-Infeld (BI) gravity theories \cite{Edelstein:2006se,Izaurieta:2009hz,Concha:2013uhq,Concha:2014vka,Concha:2014zsa}. Let us also mention that supersymmetric extensions of the Maxwell algebra found relevant applications in the development and study of supergravity models (see, for instance, \cite{Concha:2014tca,Concha:2015woa,Penafiel:2017wfr,Ravera:2018vra,Concha:2018ywv} and references therein). More recently, the dual version of the Maxwell symmetry, denoted as Hietarinta-Maxwell \cite{Bansal:2018qyz}, allowed to recover the topological and minimal massive gravity theories. In particular, both of these massive gravity theories appear as particular cases of a more general massive gravity arising from a spontaneous breaking of a local symmetry in a CS gravity theory invariant under the Hietarinta-Maxwell algebra \cite{Chernyavsky:2020fqs}.

One may then ask if the MEB gravity can admit a cosmological constant in a similar way to the one based on the Newton-Hooke symmetry \cite{Bacry:1968zf,Gibbons:2003rv,Brugues:2006yd,Alvarez:2007fw,Papageorgiou:2010ud,Duval:2011mi,Hartong:2016yrf,Duval:2016tzi}. The answer to the question is affirmative. Indeed, there is the enlarged extended Bargmann gravity (EEB) which reproduces the MEB one in the vanishing cosmological constant limit $\ell \rightarrow \infty$ \cite{Concha:2019lhn}. The EEB has been obtained as an Inönü-Wigner (IW) contraction \cite{IW, WW} of the [AdS-Lorentz] $\oplus \, \mathfrak{u}(1) \oplus \, \mathfrak{u}(1) \oplus \, \mathfrak{u}(1)$ algebra in three spacetime dimensions. The AdS-Lorentz (AdS-$\mathcal{L}$, for short) algebra has been first introduced in arbitrary spacetime dimensions as a semi-simple extension of the Poincaré algebra in \cite{Soroka:2004fj,Soroka:2006aj} and has been useful to recover pure Lovelock theory \cite{Cai:2006pq,Dadhich:2012ma} as a particular limit of CS and BI gravity theories \cite{Concha:2016kdz,Concha:2016tms,Concha:2017nca}. Moreover, supersymmetric extentions of the AdS-$\mathcal{L}$ algebra proved useful in the context of supergravity (see \cite{Concha:2015tla,Ipinza:2016con,Banaudi:2018zmh,Penafiel:2018vpe,Concha:2020atg} and references therein).

It is natural to address the question whether the extended Newtonian gravity presented in \cite{Ozdemir:2019orp} and its exotic version  admit a Maxwell generalization in three spacetime dimensions which can be formulated as a CS theory. The motivation to consider three-dimensional CS gravity theories is twofold. First, CS formalism offers us a simple framework to construct gauge invariant gravity actions. Second, three-dimensional geometry can be seen as an interesting toy model to approach more difficult theories in higher dimensions. In this paper we will show that a three-dimensional Maxwellian extended Newtonian (MENt) gravity theory can be defined by generalizing the results obtained in \cite{Bergshoeff:2020fiz}. In particular, the MENt algebra appears as a NR limit of a particular enhancement of the Maxwell gravity enlarged with three $\mathfrak{u}(1)$ gauge generators. The introduction of a cosmological constant to the MENt theory is also explored by defining a new NR gravity theory denoted as enlarged extended Newtonian (EEN) gravity which is the NR version of an enhanced AdS-$\mathcal{L}$ algebra enlarged with three $\mathfrak{u}(1)$ gauge generators.  Interestingly, we show that the MENt and EEN symmetries contain the MEB and EEB algebras as sub-cases, respectively. Nevertheless, the respective novel NR CS actions do not contain the MEB and EEB gravity actions as particular cases. This is mainly due to the problematic to define a unique rescaling for the relativistic parameters of the invariant tensor. Such difficulty can be overcame by considering the semigroup expansion ($S$-expansion) method \cite{Izaurieta:2006zz}. In particular, we show that the new NR symmetries presented here can alternatively been obtained as $S$-expansion of the enhanced Nappi-Witten algebra introduced in \cite{Bergshoeff:2020fiz}. Remarkably, the $S$-expansion procedure not only allows us to obtain the MENt and EEN algebras but also provides us with the most general non-degenerate invariant tensor for the respective NR algebras.

The organization of the paper is as follows: in Section 2, we briefly review the relativistic Maxwell and AdS-$\mathcal{L}$ gravity theories in three spacetime dimensions. In Section 3, we present new relativistic algebras enlarged with three $\mathfrak{u}(1)$ generators which can be seen as enhancements of the Maxwell and AdS-$\mathcal{L}$ symmetries. The respective CS actions based on the enhanced algebras are also constructed. In Section 4, we present the Maxwellian version of the extended Newtonian gravity. In particular, we introduce the MENt algebra as a NR limit of the [enhanced Maxwell] $\oplus\,\mathfrak{u}(1)\oplus\mathfrak{u}(1)\oplus\mathfrak{u}(1)$ algebra. The NR limit of the relativistic CS action is also considered. The generalization of our results to the presence of a cosmological constant is explored in Section 5. A novel NR symmetry denoted as EEN algebra is introduced by considering the NR limit of the [enhanced AdS-$\mathcal{L}$] $\oplus\,\mathfrak{u}(1)\oplus\mathfrak{u}(1)\oplus\mathfrak{u}(1)$ algebra. We work out that the MENt gravity theory appears as a flat limit $\ell \rightarrow \infty$ of the EEN one. In Section 6, we show that our results can alternatively be obtained by considering the $S$-expansion method. Section 7 concludes our work with some discussions about future developments.

\section{Relativistic Maxwell and AdS-Lorentz gravity theories}

In this section, we briefly review the three-dimensional relativistic Maxwell
CS gravity \cite{Salgado:2014jka,Hoseinzadeh:2014bla,Concha:2018zeb,Concha:2018jxx,Concha:2019icz} and its generalization to the so-called AdS-Lorentz gravity \cite{Diaz:2012zza,Fierro:2014lka,Concha:2018jjj}. The latter allows to include a cosmological constant into the Maxwell CS
gravity action.

The Maxwell algebra can be seen as an extension and deformation of the
Poincaré algebra and it is spanned by the set of generators $\left\{
J_{A},P_{A},Z_{A}\right\} $ which satisfy the following non-vanishing
commutation relations%
\begin{eqnarray}
\left[ J_{A},J_{B}\right] &=&\epsilon _{ABC}J^{C}\,,  \notag \\
\left[ J_{A},P_{B}\right] &=&\epsilon _{ABC}P^{C}\,,  \notag \\
\left[ J_{A},Z_{B}\right] &=&\epsilon _{ABC}Z^{C}\,,  \notag \\
\left[ P_{A},P_{B}\right] &=&\epsilon _{ABC}Z^{C}\,,  \label{Max}
\end{eqnarray}%
where $A,B,C=0,1,2$ are Lorentz indices which are lowered and raised with
the Minkowski metric $\eta _{AB}=\left( -1,1,1\right) $ and $\epsilon _{ABC}$
is the three-dimensional Levi Civita tensor which satisfy $\epsilon
_{012}=-\epsilon ^{012}=1$. Here, $J_{A}$ correspond to the spacetime
rotations, $P_{A}$ are the spacetime translations and $Z_{A}$ are the
so-called gravitational Maxwell generators.

Such relativistic algebra admits the following non-vanishing components of
the invariant tensor of rank 2,%
\begin{eqnarray}
\left\langle J_{A}J_{B}\right\rangle &=&\sigma _{0}\eta _{AB}\,,  \notag \\
\left\langle J_{A}P_{B}\right\rangle &=&\sigma _{1}\eta _{AB}\,,  \notag \\
\left\langle J_{A}Z_{B}\right\rangle &=&\sigma _{2}\eta _{AB}\,,  \notag
\\
\left\langle P_{A}P_{B}\right\rangle &=&\sigma _{2}\eta _{AB}\,,  \label{IT}
\end{eqnarray}%
where $\sigma _{0}$, $\sigma _{1}$, and $\sigma _{2}$ are arbitrary constants.

On the other hand, the gauge connection one-form $A$ for the Maxwell algebra
reads%
\begin{equation}
A=W^{A}J_{A}+E^{A}P_{A}+K^{A}Z_{A}\,,  \label{1f}
\end{equation}%
where $W^{A}$ denotes the spin-connection, $E^{A}$ is the vielbein and $%
K^{A} $ is the gravitational Maxwell gauge field. The
corresponding curvature two-form is given by%
\begin{equation}
F=R^{A}\left( W\right) J_{A}+R^{A}\left( E\right) P_{A}+R^{A}\left( K\right)
Z_{A}\,,
\end{equation}%
with%
\begin{eqnarray}
R^{A}\left( W\right) &:=& dW^{A}+\frac{1}{2}\epsilon ^{ABC}W_{B}W_{C}\,,
\notag \\
R^{A}\left( E\right) &:=& dE^{A}+\epsilon ^{ABC}W_{B}E_{C}\,,  \notag
\\
R^{A}\left( K\right) &:=& dK^{A}+\epsilon ^{ABC}W_{B}K_{C}+\frac{1}{2}%
\epsilon ^{ABC}E_{B}E_{C}\,.  \label{MaxCurv}
\end{eqnarray}%
Here as well as in the sequel, the index ``$A$'' of $W^A$, $E^A$, $K^A$ in the parenthesis on the left-hand side is understood in order to lighten the notation, meaning, in fact, $R^{A}\left( W\right) = R^{A}\left( W^B\right)$ and so on.

Considering the gauge connection one-form (\ref{1f}) and the non-vanishing
components of the invariant tensor (\ref{IT}) in the three-dimensional CS
expression,%
\begin{equation}
I=\frac{k}{4\pi }\int \left\langle AdA+\frac{2}{3}A^{3}\right\rangle \,,
\label{CS}
\end{equation}%
with $k$ being the CS level of the theory related to the gravitational
constant $G$, we find the following relativistic CS gravity action for the
Maxwell algebra \cite{Salgado:2014jka,Hoseinzadeh:2014bla,Concha:2018zeb}:
\begin{eqnarray}
I_{\text{Maxwell}} &=&\frac{k}{4\pi }\int \Bigg [ \sigma _{0}\left( W^{A}dW_{A}+%
\frac{1}{3}\epsilon ^{ABC}W_{A}W_{B}W_{C}\right) +2\sigma
_{1}E_{A}R^{A}\left( W\right) \notag \\
&& +\sigma _{2}\left( T^{A}E_{A}+2K_{A}R^{A}\left( W\right) \right) \Bigg ] \,,  \label{MaxCS}
\end{eqnarray}%
where $T^{A}=R^{A}\left( E\right) $ denotes the torsion two-form. One can
see that the CS action contains three independent sectors proportional to $%
\sigma _{0}$, $\sigma _{1}$, and $\sigma _{2}$. The gravitational
Lagrangian \cite{Witten:1988hc} appears along the $\sigma _{0}$ constant, while the Poincaré
Lagrangian given by the Einstein-Hilbert term is related to the $\sigma _{1}$
constant. On the other hand, the term proportional to $\sigma _{2}$ contains
a torsional term and the gravitational Maxwell gauge field $K^{A}$ coupled
to the gravitational gauge field. As mentioned in \cite{Concha:2018zeb}, the presence of
the additional gauge field $K^{A}$ modifies not only the asymptotic sector
but also the vacuum of the theory.

Interestingly, an enlarged relativistic algebra allows us to include a
cosmological constant into the Maxwell CS gravity theory. Such enlarged
algebra is the so-called AdS-$\mathcal{L}$ algebra $\left\{
J_{A},P_{A},Z_{A}\right\} $ whose generators satisfy (\ref{Max}) along with%
\begin{eqnarray}
\left[ Z_{A},Z_{B}\right] &=&\frac{1}{\ell ^{2}}\epsilon _{ABC}Z^{C}\,,
\notag \\
\left[ P_{A},Z_{B}\right] &=&\frac{1}{\ell ^{2}}\epsilon _{ABC}P^{C}\,,
\label{AdSL}
\end{eqnarray}%
where $\ell $ is a length parameter related to the cosmological constant $%
\Lambda $. Naturally, the flat limit $\ell \rightarrow \infty $ reproduces
the Maxwell algebra. One can see that the following redefinition of the
generators%
\begin{eqnarray}
J_{A} &=&\ell ^{2}\hat{Z}_{A}\,,  \notag \\
P_{A} &=&\hat{P}_{A}\,, \notag \\
Z_{A} &=&\hat{J}_{A}-\ell ^{2}\hat{Z}_{A}\,,
\end{eqnarray}%
allows to rewrite the algebra given by (\ref{Max}) and (\ref{AdSL}) as the
direct sum of the $\mathfrak{so}\left( 2,2\right) $ and $\mathfrak{so}\left(
2,1\right) $ algebras.

The non-vanishing components of the invariant tensor for the AdS-$\mathcal{L}$
algebra are given by (\ref{IT}) along with \cite{Concha:2018jjj}%
\begin{eqnarray}
\left\langle Z_{A}P_{B}\right\rangle &=&\frac{\sigma _{1}}{\ell ^{2}}\eta
_{AB}\,,  \notag \\
\left\langle Z_{A}Z_{B}\right\rangle &=&\frac{\sigma _{2}}{\ell ^{2}}\eta
_{AB}\,.  \label{IT2}
\end{eqnarray}%
Then, considering the gauge connection one-form for the AdS-$\mathcal{L}$ algebra
which coincides with the Maxwell one (\ref{1f}) and the invariant tensor (%
\ref{IT}) and (\ref{IT2}) into the general expression of the CS action (\ref%
{CS}), we find the following three-dimensional relativistic CS action \cite{Concha:2018jjj}:%
\begin{align}
& I_{\text{AdS-}\mathcal{L}}=\frac{k}{4\pi }\int \left[ \sigma _{0}\left(
W^{A}dW_{A}+\frac{1}{3}\,\epsilon ^{ABC}W_{A}W_{B}W_{C}\right) \right.
\notag \\
& +\left. \sigma _{1}\left( 2E_{A}R^{A}(W)+\frac{2}{\ell ^{2}}%
\,E_{A}F^{A}(K)+\frac{1}{3\ell ^{2}}\,\epsilon ^{ABC}E_{A}E_{B}E_{C} \right)
\right.  \notag \\
& +\left. \sigma _{2}\left( T^{A}E_{A}+\frac{1}{ \ell ^{2}}\,\epsilon
^{ABC}E_{A}K_{B}E_{C}+2K_{A}R^{A}(W)+\frac{1}{\ell ^{2}}K_{A}\,D_{W}K^{A}+%
\frac{1}{3\ell ^{4}}\epsilon ^{ABC}K_{A}K_{B}K_{C} \right) \right]\,,
\end{align}%
where we have defined the curvature $F^{A}(K):=D_{W}K^{A}+\frac{1}{2\ell ^{2}%
}\epsilon ^{ABC}K_{B}K_{C}$, being $D_{W}$ the Lorentz covariant
derivative $D_{W}\Theta ^{A}=d\Theta ^{A}+\epsilon ^{ABC}W_{B}\Theta _{C}.$
One can see that such relativistic symmetry not only introduces a
comological constant term to the Einstein-Hilbert term but also includes the
gauge field $K^{A}$ along the $\sigma _{1}$ constant. Furthermore, the term
along $\sigma _{2}$ is also modified by the presence of the new commutation
relations (\ref{AdSL}). Naturally, the vanishing cosmological constant
limit $\ell \rightarrow \infty $ reproduces the relativistic Maxwell CS
gravity action (\ref{MaxCS}).

NR versions of the Maxwell and the AdS-$\mathcal{L}$ algebras have been
considered in \cite{Aviles:2018jzw,Concha:2019lhn,Penafiel:2019czp,Concha:2019mxx,Concha:2020sjt} and have required to consider $U\left( 1\right) $ enlargements in
order to have well-defined finite NR CS actions with
non-degenerate bilinear forms. In particular, as shown in \cite{Aviles:2018jzw}, the
NR limit of the [Maxwell] $\oplus \, \mathfrak{u}\left( 1\right)^3 $
algebra reproduces the MEB algebra.
On the other hand, the EEB appears as a
NR limit of the [AdS-$\mathcal{L}$] $\oplus \, \mathfrak{u}\left( 1\right)^3 $ algebra \cite{Concha:2019lhn}. Such NR algebras are, as their relativistic versions, related through a flat limit.

In what follows, we study new enhancements of the Maxwell and AdS-Lorentz
algebras whose IW contraction will lead us to novel NR algebras.

\section{Enhanced Maxwell and AdS-Lorentz algebras and \texorpdfstring{$U\left( 1\right)$}{EMAL} enlargements}

In this section, we generalize the results presented in \cite{Bergshoeff:2020fiz} to enlarged
symmetries. In particular, we present new enhancements of the [Maxwell] $\oplus \, \mathfrak{u}\left( 1\right)^3 $ and [AdS-$\mathcal{L}$] $\oplus \, \mathfrak{u}\left( 1\right)^3 $ algebras whose NR limits shall be explored in the
next section. We expect to find a Maxwellian version of the Extended
Newtonian gravity \cite{Ozdemir:2019orp} and a generalization including a cosmological constant.
The explicit CS action for the relativistic enhancement symmetries is also
constructed.

\subsection{Enhanced Maxwell \texorpdfstring{$\oplus \, \mathfrak{u}\left( 1\right) \, \oplus \, \mathfrak{u}\left( 1\right) \, \oplus \, \mathfrak{u}\left( 1\right) $}{EMu1} algebra}

A particular extension of the Maxwell algebra can be obtained by adding the
set of generators $\left\{ S_{A},T_{A},V_{A}\right\} $ to the usual Maxwell
ones $\left\{ J_{A},P_{A},Z_{A}\right\} $. Such extension satisfies the
following non-vanishing commutation relations%
\begin{eqnarray}
\left[ J_{A},J_{B}\right] &=&\epsilon _{ABC}J^{C}\,,\qquad \left[ J_{A},S_{B}%
\right] =\epsilon _{ABC}S^{C}\,,  \notag \\
\left[ J_{A},P_{B}\right] &=&\epsilon _{ABC}P^{C}\,,\qquad \left[ J_{A},T_{B}%
\right] =\epsilon _{ABC}T^{C}\,,  \notag \\
\left[ J_{A},Z_{B}\right] &=&\epsilon _{ABC}Z^{C}\,,\qquad \left[ S_{A},P_{B}%
\right] =\epsilon _{ABC}T^{C}\,, \notag \\
\left[ P_{A},P_{B}\right] &=&\epsilon _{ABC}Z^{C}\,,\qquad \left[ J_{A},V_{B}%
\right] =\epsilon _{ABC}V^{C}\,,  \notag \\
\left[ S_{A},Z_{B}\right] &=&\epsilon _{ABC}V^{C}\,,\qquad \left[ T_{A},P_{B}%
\right] =\epsilon _{ABC}V^{C}\,.\label{EMax}
\end{eqnarray}%
Such algebra can be seen as an extension and deformation of the coadjoint
Poincaré algebra studied in \cite{Barducci:2019jhj,Bergshoeff:2020fiz,Barducci:2020blv}. Interestingly, the commutation relations (\ref{EMax}) can also be recovered
as an semigroup expansion ($S$-expansion) \cite{Izaurieta:2006zz} of the Poincaré algebra $%
\mathfrak{iso}(2,1)=\left\{ \hat{J}_{A},\hat{P}_{A}\right\} $, whose generators
satisfy the non-vanishing commutation relations
\begin{eqnarray}
\left[ \hat{J}_{A},\hat{J}_{B}\right] &=&\epsilon _{ABC}\hat{J}^{C}\,,
\notag \\
\left[ \hat{J}_{A},\hat{P}_{B}\right] &=&\epsilon _{ABC}\hat{P}^{C}\,.
\label{eL}
\end{eqnarray}%
Indeed, let us consider $S_{E}^{\left( 2\right) }=\left\{ \lambda
_{0},\lambda _{1},\lambda _{2},\lambda _{3}\right\} $ as the relevant
semigroup whose elements satisfy%
\begin{equation}
\lambda _{\alpha }\lambda _{\beta }=\left\{
\begin{array}{lcl}
\lambda _{\alpha +\beta }\,\,\,\, & \mathrm{if}\,\,\,\,\alpha +\beta \leq 3 \, ,
&  \\
\lambda _{3}\,\,\, & \mathrm{if}\,\,\,\,\alpha +\beta >3 \, , &
\end{array}%
\right.  \label{ml}
\end{equation}%
where $\lambda _{3}=0_{s}$ is the zero element of the semigroup such that $%
0_{s}\lambda _{\alpha }=0_{s}$. Then, one can show that the enhanced Maxwell
algebra (\ref{EMax}) appears as a $0_{s}$-reduction of the expanded algebra $S_{E}^{\left( 2\right)} \times \mathfrak{iso}(2,1)$, where the expanded generators are
expressed in terms of the Poincaré ones through the semigroup elements as%
\begin{eqnarray}
J_{A} &=&\lambda _{0}\hat{J}_{A}\,,\qquad S_{A}=\lambda _{0}\hat{P}_{A}\,,
\notag \\
P_{A} &=&\lambda _{1}\hat{J}_{A}\,,\qquad T_{A}=\lambda _{1}\hat{P}_{A}\,,
\notag \\
Z_{A} &=&\lambda _{2}\hat{J}_{A}\,,\qquad V_{A}=\lambda _{2}\hat{P}_{A}\,.
\label{Exp}
\end{eqnarray}
Let us note that the possibility to obtain generalizations of the coadjoint Poincaré algebra by considering $S_{E}^{(N)}$-expansion of the $\mathfrak{iso}(2,1)$ algebra has first been discussed in \cite{Bergshoeff:2020fiz}. In the present analysis, we have shown that considering $N=2$ allows to define an enhanced Maxwell algebra. As we shall see in the next section, by considering a different semigroup, a new relativistic symmetry will be obtained.

An interesting feature of the $S$-expansion method is that it immediately provides us
with the non-vanishing components of the invariant tensor of the expanded
algebra in terms of the original ones \cite{Izaurieta:2006zz}. In particular, the non-vanishing
components of the invariant tensor for the Poincaré algebra are given by%
\begin{eqnarray}
\left\langle \hat{J}_{A}\hat{J}_{B}\right\rangle &=&\nu _{0}\eta _{AB}\,,
\notag \\
\left\langle \hat{J}_{A}\hat{P}_{B}\right\rangle &=&\nu _{1}\eta _{AB}\,.
\end{eqnarray}%
Thus, one can show that the enhanced Maxwell algebra admits the following
invariant tensor:
\begin{eqnarray}
\left\langle J_{A}J_{B}\right\rangle &=&\mu _{0}\eta _{AB}\,,  \notag \\
\left\langle J_{A}S_{B}\right\rangle &=&\mu _{1}\eta _{AB}\,,  \notag \\
\left\langle J_{A}P_{B}\right\rangle &=&\mu _{2}\eta _{AB}\,,\qquad  \notag
\\
\left\langle J_{A}T_{B}\right\rangle &=&\left\langle S_{A}P_{B}\right\rangle
=\mu _{3}\eta _{AB}\,,  \notag \\
\left\langle J_{A}Z_{B}\right\rangle &=&\left\langle P_{A}P_{B}\right\rangle
=\mu _{4}\eta _{AB}\,,  \notag \\
\left\langle J_{A}V_{B}\right\rangle &=&\left\langle S_{A}Z_{B}\right\rangle
=\left\langle T_{A}P_{B}\right\rangle =\mu _{5}\eta _{AB}\,,  \label{Emaxinvt}
\end{eqnarray}%
where $\mu $'s are arbitrary constants and are related to the Poincaré
constant as%
\begin{eqnarray}
\mu _{0} &=&\lambda _{0}\nu _{0}\,,\quad \mu _{1}=\lambda _{0}\nu
_{1}\,,\quad \mu _{2}=\lambda _{1}\nu _{0}\,,  \notag \\
\mu _{3} &=&\lambda _{1}\nu _{1}\,,\quad \mu _{4}=\lambda _{2}\nu
_{0}\,,\quad \mu _{5}=\lambda _{2}\nu _{1}\,.
\end{eqnarray}

Similarly to what happens in the MEB algebra, one can generalize the
relativistic algebra by including three $U\left( 1\right) $
generators given by $Y_{1}$, $Y_{2}$, and $Y_{3}$. The inclusion of these $U\left(
1\right) $ generators in the case of the original Maxwell algebra assures to have a well-defined non-degenerate invariant tensor after applying the
NR limit. Although these additional $U\left(
1\right) $ generators act as central charges in the relativistic algebra,
their presence implies additional components of the invariant tensor given
by \cite{Aviles:2018jzw}%
\begin{eqnarray}
\left\langle Y_{1}Y_{1}\right\rangle &=&\mu _{0}\,,  \notag \\
\left\langle Y_{1}Y_{2}\right\rangle &=&\mu _{2}\,,  \notag \\
\left\langle Y_{1}Y_{3}\right\rangle &=&\left\langle Y_{2}Y_{2}\right\rangle
=\mu _{4}\,.  \label{U1invt}
\end{eqnarray}%
The non-vanishing components of the invariant tensor (\ref{Emaxinvt}) and (%
\ref{U1invt}) define an invariant tensor for the [enhanced Maxwell] $\oplus \,
\mathfrak{u}\left( 1\right)^3 $ algebra.

The gauge connection one-form for the [enhanced Maxwell] $\oplus \,
\mathfrak{u}\left( 1\right) ^3 $ algebra now includes not only three additional bosonic gauge
fields, $\Sigma ^{A},L^{A}$ and $\Gamma ^{A}$ in (\ref{1f}) but also three
extra $U\left( 1\right) $ gauge field one-forms as%
\begin{equation}
A=W^{A}J_{A}+E^{A}P_{A}+K^{A}Z_{A}+\Sigma ^{A}S_{A}+L^{A}T_{A}+\Gamma
^{A}V_{A}+SY_{1}+MY_{2}+TY_{3}\,.\,  \label{e1f}
\end{equation}%
The corresponding curvature two-form is given by%
\begin{eqnarray*}
F &=&R^{A}\left( W\right) J_{A}+R^{A}\left( E\right) P_{A}+R^{A}\left(
K\right) Z_{A}+R^{A}\left( \Sigma \right) S_{A}+R^{A}\left( L\right)
T_{A}+R^{A}\left( \Gamma \right) V_{A} \\
&&+R\left( S\right) Y_{1}+R\left( M\right) Y_{2}+R\left( T\right) Y_{3}\,,
\end{eqnarray*}%
where $R^{A}\left( W\right) $, $R^{A}\left( E\right) $ and $R^{A}\left(
K\right) $ are given by (\ref{MaxCurv}) and%
\begin{eqnarray}
R^{A}\left( \Sigma \right)  &=&d\Sigma ^{A}+\epsilon ^{ABC}W_{B}\Sigma
_{C}\,,  \notag \\
R^{A}\left( L\right)  &=&dL^{A}+\epsilon ^{ABC}W_{B}L_{C}+\epsilon
^{ABC}\Sigma _{B}E_{C}\,,  \notag \\
R^{A}\left( \Gamma \right)  &=&d\Gamma ^{A}+\epsilon ^{ABC}W_{B}\Gamma
_{C}+\epsilon ^{ABC}\Sigma _{B}K_{C}+\epsilon ^{ABC}L_{B}E_{C}\,.  \label{EMaxCurv}
\end{eqnarray}%
Then, considering the gauge connection one-form (\ref{e1f}) and the
non-vanishing components of the invariant tensor (\ref{Emaxinvt}) and (\ref%
{U1invt}) in the general expression of the CS action (\ref{CS}), we find the
following relativistic CS action for the [enhanced Maxwell] $\oplus \, \mathfrak{u%
}\left( 1\right) ^{3}$ algebra:
\begin{eqnarray}
I_{\text{enh-Maxwell} \, \oplus \, \mathfrak{u}\left( 1\right) ^{3}} &=&\frac{k}{4\pi }%
\int \Bigg [ \mu _{0}\left( W^{A}dW_{A}+\frac{1}{3}\epsilon
^{ABC}W_{A}W_{B}W_{C} + SdS\right) +2\mu _{1}\Sigma _{A}R^{A}\left( W\right)
   \notag \\
&&+  2\mu _{2}\left( E_{A}R^{A}\left( W\right) + MdS\right) +2\mu
_{3}\left( L_{A}R^{A}\left( W\right) +E_{A}R^{A}\left( \Sigma \right)
\right)    \notag \\
&&+  \mu _{4}\left( T^{A}E_{A}+2K_{A}R^{A}\left( W\right)
+MdM+2SdT\right)   \notag \\
&&+ 2\mu _{5}\left( \Gamma _{A}R^{A}\left( W\right) +E_{A}R^{A}\left(
L\right) +K_{A}R^{A}\left( \Sigma \right)-\frac{1}{2}\epsilon^{ABC}\Sigma_{A}E_{B}E_{C} \right) \Bigg ] \,, \notag \\ \label{EMaxCS}
\end{eqnarray}%
where $T^{A}=D_{W}E^{A}$. It is interesting to note that the terms proportional to $\mu _{2}$ and $\mu
_{3}$ reproduce the CS action for the coadjoint Poincaré algebra presented
in \cite{Bergshoeff:2020fiz}. In particular, the Einstein-Hilbert term appears along $\mu _{2}$.
On the other hand, the term $\mu _{0}$ contains the usual exotic Lagrangian \cite{Witten:1988hc}. The terms proportional to $\mu _{1},\mu _{3}$, and $\mu _{5}$ contains
new contributions to the Maxwell CS gravity action given by the additional
set of gauge fields. The
non-degeneracy of the invariant tensor implies that the gauge fields are
dynamically determined by the vanishing of every curvature (\ref{MaxCurv})
and (\ref{EMaxCurv}) provided $\mu _{5}\neq 0$. In particular, the vanishing
of $R_{A}\left( K\right) $ can be seen as the constancy of the background
electromagnetic field in flat spacetime.

As was discussed in \cite{Aviles:2018jzw}, the inclusion of three extra $%
U\left( 1\right) $ gauge fields to the Maxwell algebra allows to reproduce a non-degenerate invariant tensor at the NR level. Here, we shall see that the\ addition of
those $U\left( 1\right) $ gauge fields in the enhanced Maxwell algebra (\ref%
{EMax}) will be crucial to obtain a well-defined Maxwellian version of the
extended Newtonian gravity.

\subsection{Enhanced AdS-Lorentz \texorpdfstring{$\oplus \, \mathfrak{u}\left( 1\right) \, \oplus \, \mathfrak{u}\left( 1\right) \, \oplus \, \mathfrak{u}\left( 1\right) $}{EALu1} algebra}

A cosmological constant term can be introduced to the [enhanced Maxwell] $\oplus \,
\mathfrak{u}\left( 1\right)^3 $ gravity
theory by enlarging the algebra and introducing a length scale $\ell $. The
new algebra can be seen as an extension of the AdS-$\mathcal{L}$ algebra (\ref{Max}%
)-(\ref{AdSL}) by adding the set of generators $\left\{
S_{A},T_{A},V_{A}\right\} $. Such extension satisfies the non-vanishing
commutation relations (\ref{EMax}) along with%
\begin{eqnarray}
\left[ Z_{A},Z_{B}\right] &=&\frac{1}{\ell ^{2}}\epsilon
_{ABC}Z^{C}\,,\qquad \left[ T_{A},Z_{B}\right] =\frac{1}{\ell ^{2}}\epsilon
_{ABC}T^{C}\,,  \notag \\
\left[ P_{A},Z_{B}\right] &=&\frac{1}{\ell ^{2}}\epsilon
_{ABC}P^{C}\,,\qquad \left[ V_{A},P_{B}\right] =\frac{1}{\ell ^{2}}\epsilon
_{ABC}T^{C}\,,  \notag \\
\left[ V_{A},Z_{B}\right] &=&\frac{1}{\ell ^{2}}\epsilon _{ABC}V^{C}\,.
\label{EAdSL}
\end{eqnarray}%
It is interesting to note that such enhanced AdS-$\mathcal{L}$ algebra can be
rewritten as three copies of the Poincaré algebra, that is to say
\begin{eqnarray}
\left[ J_{A}^{\pm },J_{B}^{\pm }\right] &=&\epsilon _{ABC}J^{\pm C}\,,\qquad %
\left[ \hat{J}_{A},\hat{J}_{B}\right] =\epsilon _{ABC}\hat{J}^{C}\,,  \notag
\\
\left[ J_{A}^{\pm },P_{B}^{\pm }\right] &=&\epsilon _{ABC}P^{\pm C}\,,\qquad %
\left[ \hat{J}_{A},\hat{P}_{B}\right] =\epsilon _{ABC}\hat{P}^{C}\,,
\label{3iso}
\end{eqnarray}%
by considering the following redefinition of the generators:
\begin{eqnarray}
J_{A} &=&\hat{J}_{A}+J_{A}^{+}+J_{A}^{-}\,,\qquad S_{A}=\hat{P}%
_{A}+P_{A}^{+}+P_{A}^{-}\,,  \notag \\
P_{A} &=&\frac{1}{\ell }\left( J_{A}^{+}-J_{A}^{-}\right) \,,\qquad \ T_{A}=%
\frac{1}{\ell }\left( P_{A}^{+}-P_{A}^{-}\right) \,, \notag \\
Z_{A} &=&\frac{1}{\ell ^{2}}\left( J_{A}^{+}+J_{A}^{-}\right) \,,\qquad
V_{A}=\frac{1}{\ell ^{2}}\left( P_{A}^{+}+P_{A}^{-}\right) \,.
\end{eqnarray}%
The relativistic enhanced AdS-$\mathcal{L}$ algebra can also be written as the direct sum of the
coadjoint AdS algebra $\left\{ \tilde{J}_{A},\tilde{P}_{A},\tilde{S}_{A},%
\tilde{T}_{A}\right\} $ defined in \cite{Bergshoeff:2020fiz} and the Poincaré algebra $\left\{
\check{J}_{A},\check{P}_{A}\right\} $ by considering the following
redefinition:
\begin{eqnarray}
J_{A} &=&\tilde{J}_{A}+\check{J}_{A}\,,\qquad S_{A}=\tilde{S}_{A}+\check{P}%
_{A}\,,  \notag \\
P_{A} &=&\tilde{P}_{A}\,,\qquad \qquad \ T_{A}=\tilde{T}_{A}\,, \notag \\
Z_{A} &=&\frac{\tilde{J}_{A}}{\ell ^{2}}\,,\qquad \quad \ \ \ \, V_{A}=\frac{%
\tilde{S}_{A}}{\ell ^{2}}\,.
\end{eqnarray}%
Although the relativistic algebra seems simpler in the form of three copies
of the Poincaré algebra (\ref{3iso}), we shall consider the basis $\left\{
J_{A},P_{A},Z_{A},S_{A},T_{A},V_{A}\right\} $ since it allows to establish a
well-defined and evident vanishing cosmological constant limit $\ell \rightarrow \infty $
reproducing the enhanced Maxwell algebra. As we shall see, the same behavior
will be inherit to its NR version.

An alternative procedure to obtain the enhanced AdS-$\mathcal{L}$ algebra ((\ref%
{EMax}) and (\ref{EAdSL})) is given by the $S$-expansion method \cite{Izaurieta:2006zz}. Indeed,
let us consider $S_{\mathcal{M}}^{\left( 2\right) }=\left\{ \lambda
_{0},\lambda _{1},\lambda _{2}\right\} $ as the relevant semigroup and the
Poincaré algebra $\mathfrak{iso}(2,1)$ as the original one. The semigroup elements satisfy the
following multiplication law:%
\begin{equation}
\lambda _{\alpha }\lambda _{\beta }=\left\{
\begin{array}{lcl}
\lambda _{\alpha +\beta }\,\,\,\, & \mathrm{if}\,\,\,\,\alpha +\beta \leq 2 \, ,
&  \\
\lambda _{\alpha +\beta -2}\,\,\, & \mathrm{if}\,\,\,\,\alpha +\beta >2 \, . &
\end{array}%
\right.  \label{ml2}
\end{equation}%
Then, one can show that the enhanced AdS-$\mathcal{L}$ algebra given by (\ref{EMax}%
) and (\ref{EAdSL}) appears as a $S_{\mathcal{M}}^{\left( 2\right) }$%
-expansion of the Poincaré algebra where the expanded generators are
expressed in terms of the Poincaré ones through the semigroup elements as%
\begin{eqnarray}
J_{A} &=&\lambda _{0}\hat{J}_{A}\,,\qquad \ \ S_{A}=\lambda _{0}\hat{P}%
_{A}\,,  \notag \\
\ell P_{A} &=&\lambda _{1}\hat{J}_{A}\,,\qquad \ \ell T_{A}=\lambda _{1}\hat{%
P}_{A}\,, \notag \\
\ell ^{2}Z_{A} &=&\lambda _{2}\hat{J}_{A}\,,\qquad \ell ^{2}V_{A}=\lambda
_{2}\hat{P}_{A}\,.
\end{eqnarray}%
Then, following theorem VII.2 of \cite{Izaurieta:2006zz}, one can show that the non-vanishing
components of the enhanced AdS-$\mathcal{L}$ algebra can be expressed in terms of
the Poincaré ones. Thus, the invariant tensor for the present relativistic
algebra are given by (\ref{Emaxinvt}) along with%
\begin{eqnarray}
\left\langle Z_{A}P_{B}\right\rangle &=&\frac{\mu _{2}}{\ell ^{2}}\eta
_{AB}\,,  \notag \\
\left\langle T_{A}Z_{B}\right\rangle &=&\left\langle V_{A}P_{B}\right\rangle
=\frac{\mu _{3}}{\ell ^{2}}\eta _{AB}\,,  \notag \\
\left\langle Z_{A}Z_{B}\right\rangle &=&\frac{\mu _{4}}{\ell ^{2}}\eta
_{AB}\,,  \notag \\
\left\langle V_{A}Z_{B}\right\rangle &=&\frac{\mu _{5}}{\ell ^{2}}\eta
_{AB}\,.  \label{EAdSLinvt}
\end{eqnarray}%
Naturally, the flat limit $\ell \rightarrow \infty $ reproduces the
invariant tensor of the enhanced Maxwell one. Analogously to the procedure
to obtain the EEB algebra \cite{Concha:2019lhn}, one can generalize the relativistic algebra
by including three $\mathfrak{u}\left( 1\right) $ generators given by $Y_{1}$%
, $Y_{2}$, and $Y_{3}$ which provide the following non-vanishing components of the invariant tensor \cite{Concha:2019lhn}:
\begin{eqnarray}
\left\langle Y_{1}Y_{1}\right\rangle &=&\mu _{0}\,,\qquad \left\langle
Y_{2}Y_{2}\right\rangle =\mu _{4}\,,  \notag \\
\left\langle Y_{1}Y_{2}\right\rangle &=&\mu _{2}\,,\qquad \left\langle
Y_{2}Y_{3}\right\rangle =\frac{\mu _{2}}{\ell ^{2}}\,,  \notag \\
\left\langle Y_{1}Y_{3}\right\rangle &=&\,\mu _{4}\,,\qquad \left\langle
Y_{3}Y_{3}\right\rangle =\frac{\mu _{4}}{\ell ^{2}}\,.  \label{U1invtb}
\end{eqnarray}%
The invariant tensor given by (\ref{Emaxinvt}), (\ref{EAdSLinvt}), and (\ref{U1invtb}) defines a non-degenerate invariant bilinear form for the [enhanced AdS-$\mathcal{L}$] $\oplus \, \mathfrak{u}\left( 1\right)^3 $ algebra.

Then, considering the gauge connection one-form for the enhanced algebra which coincides with the
Maxwell one (\ref{e1f}) and the non-vanishing components of the invariant
tensor (\ref{Emaxinvt}), (\ref{EAdSLinvt}), and (\ref{U1invt}) in the general
expression of the CS action (\ref{CS}), we find the following relativistic
CS action for the [enhanced AdS-$\mathcal{L}$] $\oplus \, \mathfrak{u}\left( 1\right)
^{3}$ algebra:
\begin{eqnarray}
I_{\text{enh-AdS-}\mathcal{L}\,\oplus \,\mathfrak{u}\left( 1\right) ^{3}}
&=&I_{\text{enh-Maxwell}\,\oplus \,\mathfrak{u}\left( 1\right) ^{3}}+\frac{k%
}{4\pi \ell ^{2}}\int \left[ 2\mu _{2}\left( E_{A}F^{A}\left( K\right) +%
\frac{1}{6}\epsilon ^{ABC}E_{A}E_{B}E_{C}+MdT\right) \right.   \notag \\
&&+\left. 2\mu _{3}\left( L_{A}F^{A}\left( K\right) -\frac{1}{2}\epsilon
^{ABC}E_{A}E_{B}L_{C}+E_{A} R^{A}\left( \Gamma \right) +\frac{1}{\ell^{2}}\epsilon
^{ABC}E_{A}K_{B}\Gamma _{C}\right) \right.   \notag \\
&&+\left. \mu _{4}\left( \epsilon ^{ABC}E_{A}K_{B}E_{C}+ K_{A}D_{W}K^{A}+%
\frac{1}{3 \ell^{2}}\epsilon ^{ABC}K_{A}K_{B}K_{C}+TdT\right) \right. \notag \\
&&+\left. 2\mu _{5}\left( \Gamma _{A}F^{A}\left( K\right) +\frac{1}{2}%
\epsilon ^{ABC}E_{A}E_{B}\Gamma _{C}+\frac{1}{2}\epsilon^{ABC}K_{A}\Sigma_{B}K_{C}\right. \right.   \notag \\
&&+\left. \left. \epsilon^{ABC}K_{A}L_{B}E_{C}\right) \right] \,, \label{EAdSLCS}
\end{eqnarray}%
where
\begin{eqnarray}
R^{A}\left( \Gamma \right)  &=&d\Gamma ^{A}+\epsilon ^{ABC}W_{B}\Gamma
_{C}+\epsilon ^{ABC}\Sigma _{B}K_{C}+\epsilon ^{ABC}L_{B}E_{C}\,,\nonumber\\
F^{A}\left(K\right)&=&D_{W}K^{A}+\frac{1}{2\ell ^{2}}\epsilon ^{ABC}K_{B}K_{C}\,.
\end{eqnarray}
One can see that the terms proportional to $\mu _{2}$, $\mu _{3}$, $\mu _{4}$, and $\mu _{5}$ contain new contributions from the new commutation relations (\ref{EAdSL})
of the [enhanced AdS-$\mathcal{L}$] $\oplus \,
\mathfrak{u}\left( 1\right)^3 $ algebra. In
particular, a cosmological constant term is added to the $\mu _{2}$ term
together with the gauge field $K^{A}$. Interestingly, the flat limit $\ell
\rightarrow \infty $ reproduces the relativistic [enhanced Maxwell] $\oplus \,
\mathfrak{u}\left( 1\right) ^{3}$ gravity action. As we shall see, the
presence of the $U\left( 1\right) $ gauge fields will be essential to establish a
well-defined NR limit allowing us to accommodate a cosmological
constant into the Maxwellian version of the extended Newtonian gravity.

\section{Maxwellian extended Newtonian gravity}

In this section, we apply a NR limit to the previously introduced [enhanced Maxwell] $\oplus \, \mathfrak{u}\left( 1\right)^3 $ gravity theory in three spacetime dimensions. We show that the new
NR algebra corresponds to a Maxwellian version of the extended
Newtonian algebra introduced in \cite{Ozdemir:2019orp} and subsequently studied in \cite{Concha:2019dqs,Bergshoeff:2020fiz}. The three-dimensional NR CS action
based on this new symmetry is also discussed.

\subsection{Maxwellian extended Newtonian algebra and non-relativistic limit}

A NR version of the [enhanced Maxwell] $\oplus \,
\mathfrak{u}(1)^{3}$ algebra can be obtained through an IW contraction. To this
end, we consider a dimensionless parameter $\xi $ and we express the
relativistic enhanced Maxwell generators as a linear combination of the
NR ones (denoted with a tilde) as%
\begin{eqnarray}
J_{0} &=&\frac{\tilde{J}}{2} + \xi^2 \tilde{S} -\xi ^{4}\tilde{B}\,,\qquad \quad \ \ J_{a}=\frac{%
\xi }{2}\tilde{G}_{a}-\frac{\xi ^{3}}{2}\tilde{B}_{a}\,,  \notag \\
P_{0} &=&\frac{\tilde{H}}{2} + \xi^2 \tilde{M} -\xi ^{4}\tilde{Y}\,,\qquad \quad P_{a}=\frac{%
\xi }{2}\tilde{P}_{a}-\frac{\xi ^{3}}{2}\tilde{T}_{a}\,,  \notag \\
Z_{0} &=&\frac{\tilde{Z}}{2} + \xi^2 \tilde{T} -\xi ^{4}\tilde{W}\,,\qquad \quad \  Z_{a}=\frac{%
\xi }{2}\tilde{Z}_{a}-\frac{\xi ^{3}}{2}\tilde{V}_{a}\,,  \notag \\
S_{0} &=&-\xi ^{2}\tilde{S}\,,\qquad \qquad \qquad \qquad \, S_{a}=-\xi \tilde{G}%
_{a}-\xi ^{3}\tilde{B}_{a}\,,  \notag \\
T_{0} &=&-\xi ^{2}\tilde{M}\,,\qquad \qquad \qquad \quad \ \ \, T_{a}=-\xi \tilde{P}%
_{a}-\xi ^{3}\tilde{T}_{a}\,,  \notag \\
V_{0} &=&-\xi ^{2}\tilde{T}\,,\qquad \qquad \qquad \qquad \, V_{a}=-\xi \tilde{Z}%
_{a}-\xi ^{3}\tilde{V}_{a}\,.  \label{NR1}
\end{eqnarray}%
On the other hand, one can express the $\mathfrak{u}\left( 1\right) $
generators $Y_{1}$, $Y_{2}$ and $Y_{3}$ in terms of the NR generators as%
\begin{equation}
Y_{1}=\frac{\tilde{J}}{2}- \xi^2 \tilde{S} +\xi ^{4}\tilde{B}\,,\qquad \quad Y_{2}=\frac{\tilde{H}}{2} - \xi^2 \tilde{M}+\xi ^{4}\tilde{Y}\,,\qquad \quad Y_{3}=\frac{\tilde{Z}}{2} - \xi^2 \tilde{T}+\xi ^{4}\tilde{W} \,.
\label{NR2}
\end{equation}

After considering the contraction of the [enhanced Maxwell] $\oplus \,
\mathfrak{u}(1)^{3}$ algebra (\ref%
{EMax}) and the limit $\xi \rightarrow \infty $, we find that the generators
of the novel NR algebra satisfy the MEB algebra \cite{Aviles:2018jzw},%
\begin{eqnarray}
\left[ \tilde{J},\tilde{G}_{a}\right] &=&\epsilon _{ab}\tilde{G}_{b}\,, \ %
\left[ \tilde{G}_{a},\tilde{G}_{b}\right] =-\epsilon _{ab}\tilde{S}\,, \ %
\left[ \tilde{H},\tilde{G}_{a}\right] =\epsilon _{ab}\tilde{P}_{b}\,,  \notag
\\
\left[ \tilde{J},\tilde{P}_{a}\right] &=&\epsilon _{ab}\tilde{P}_{b}\,,\ \, %
\left[ \tilde{G}_{a},\tilde{P}_{b}\right] =-\epsilon _{ab}\tilde{M}\,,\,%
\left[ \tilde{H},\tilde{P}_{a}\right] =\epsilon _{ab}\tilde{Z}_{b}\,,  \notag
\\
\left[ \tilde{J},\tilde{Z}_{a}\right] &=&\epsilon _{ab}\tilde{Z}_{b}\,,\ %
\left[ \tilde{G}_{a},\tilde{Z}_{b}\right] =-\epsilon _{ab}\tilde{T}\,,\ \, %
\left[ \tilde{Z},\tilde{G}_{a}\right] =\epsilon _{ab}\tilde{Z}_{b}\,,
\notag \\
\,\left[ \tilde{P}_{a},\tilde{P}_{b}\right] &=&-\epsilon _{ab}\tilde{T}\text{%
\thinspace },  \label{MEB}
\end{eqnarray}%
along with%
\begin{eqnarray}
\left[ \tilde{J},\tilde{B}_{a}\right] &=&\epsilon _{ab}\tilde{B}_{b}\,, \ %
\left[ \tilde{G}_{a},\tilde{B}_{b}\right] =-\epsilon _{ab}\tilde{B}\,, \ %
\left[ \tilde{H},\tilde{B}_{a}\right] =\epsilon _{ab}\tilde{T}_{b}\,,  \notag
\\
\,\,\left[ \tilde{J},\tilde{T}_{a}\right] &=&\epsilon _{ab}\tilde{T}_{b}\,,\ \ %
\left[ \tilde{G}_{a},\tilde{T}_{b}\right] =-\epsilon _{ab}\tilde{Y}\,,\ \, %
\left[ \tilde{H},\tilde{T}_{a}\right] =\epsilon _{ab}\tilde{V}_{b}\,,  \notag
\\
\left[ \tilde{J},\tilde{V}_{a}\right] &=&\epsilon _{ab}\tilde{V}_{b}\,,\ \ %
\left[ \tilde{G}_{a},\tilde{V}_{b}\right] =-\epsilon _{ab}\tilde{W}\,,\,%
\left[ \tilde{Z},\tilde{B}_{a}\right] =\epsilon _{ab}\tilde{V}_{b}\,,  \notag
\\
\left[ \tilde{S},\tilde{G}_{a}\right] &=&\epsilon _{ab}\tilde{B}_{b}\,,\,\,%
\left[ \tilde{P}_{a},\tilde{B}_{a}\right] =-\epsilon _{ab}\tilde{Y}\,, \, \left[
\tilde{M},\tilde{G}_{a}\right] =\epsilon _{ab}\tilde{T}_{b}\,,  \notag
\\
\left[ \tilde{S},\tilde{P}_{a}\right] &=&\epsilon _{ab}\tilde{T}_{b}\,,\,\,\,\,%
\left[ \tilde{P}_{a},\tilde{T}_{b}\right] =-\epsilon _{ab}\tilde{W}\,,%
\left[ \tilde{M},\tilde{P}_{a}\right] =\epsilon _{ab}\tilde{V}_{b}\,,  \notag
\\
\,\,\left[ \tilde{S},\tilde{Z}_{a}\right] &=&\epsilon _{ab}\tilde{V}%
_{b}\,,\,\,\,\left[ \tilde{Z}_{a},\tilde{B}_{b}\right] =-\epsilon _{ab}%
\tilde{W}\,,\,\left[ \tilde{T},\tilde{G}_{a}\right] =\epsilon _{ab}\tilde{V}%
_{b}\,,  \label{MENt}
\end{eqnarray}%
where $a=1,2$, $\epsilon _{ab}\equiv \epsilon _{0ab}$, $\epsilon ^{ab}\equiv
\epsilon ^{0ab}$.

The new NR algebra corresponds to a Maxwellian version of the so-called extended Newtonian algebra \cite{Ozdemir:2019orp} which we have denoted as MENt algebra.
In particular, the MENt algebra contains the MEB
generators $\left\{ \tilde{J},\tilde{G}_{a},\tilde{S},\tilde{H},\tilde{P}%
_{a},\tilde{M},\tilde{Z},\tilde{Z}_{a},\tilde{T}\right\} $ together with a set of additional generators $\left\{ \tilde{T}_{a},\tilde{B}_{a},\tilde{V}_{a}, \tilde{B},\tilde{Y},\tilde{W}\right\}$. As we shall see, the three central charges $\tilde{B}$, $\tilde{Y}$, and $\tilde{W}$, appearing in the MENt algebra as a result of the NR limit, allows to have a non-degenerate bilinear form, thus assuring
the proper construction of a three-dimensional NR CS gravity action.

\subsection{Non-relativistic MENt Chern-Simons gravity action}

The construction of a three-dimensional CS action based on this new
NR symmetry requires the non-vanishing components of the
invariant tensor for the MENt algebra together with the gauge connection
one-form $A=A_{\mu }dx^{\mu }$.

The MENt algebra can be equipped with the most general extended Newtonian
non-vanishing components of the invariant tensor \cite{Concha:2019dqs},
\begin{eqnarray}
\left\langle \tilde{S}\tilde{S}\right\rangle &=&\left\langle \tilde{J}\tilde{%
B}\right\rangle =-\beta _{0}\,,  \notag \\
\left\langle \tilde{G}_{a}\tilde{B}_{b}\right\rangle &=&\beta _{0}\delta
_{ab}\,,  \notag \\
\left\langle \tilde{M}\tilde{S}\right\rangle &=&\left\langle \tilde{H}\tilde{%
B}\right\rangle =\left\langle \tilde{J}\tilde{Y}\right\rangle =-\beta _{1}\,, \notag\\
\left\langle \tilde{P}_{a}\tilde{B}_{b}\right\rangle &=&\left\langle \tilde{G%
}_{a}\tilde{T}_{b}\right\rangle =\beta _{1}\delta _{ab}\,,  \label{ENinvt}
\end{eqnarray}%
along with
\begin{eqnarray}
\left\langle \tilde{M}\tilde{M}\right\rangle &=&\left\langle \tilde{T}\tilde{%
S}\right\rangle =\left\langle \tilde{Z}\tilde{B}\right\rangle =\left\langle
\tilde{H}\tilde{Y}\right\rangle =\left\langle \tilde{J}\tilde{W}%
\right\rangle =-\beta _{2}\,,  \notag \\
\left\langle \tilde{P}_{a}\tilde{T}_{b}\right\rangle &=&\left\langle \tilde{G%
}_{a}\tilde{V}_{b}\right\rangle =\left\langle \tilde{Z}_{a}\tilde{B}%
_{b}\right\rangle =\beta _{2}\delta _{ab}\,,  \label{MENtinvt}
\end{eqnarray}%
where the relativistic parameters $\mu $'s were rescaled as%
\begin{equation}\label{betasMENt}
\mu _{0}=\mu _{1}=-\beta _{0}\xi ^{4}\,,\qquad \mu _{2}=\mu _{3}=-\beta
_{1}\xi ^{4}\,,\qquad \mu _{4}=\mu _{5}=-\beta _{2}\xi ^{4}\,.
\end{equation}%
In particular, $\beta _{0}$ is related to an exotic sector of the extended
Newtonian gravity \cite{Concha:2019dqs}. Furthermore, the MENt algebra also admits the MEB
non-vanishing components of the invariant tensor given by \cite{Aviles:2018jzw} (see also \cite{Concha:2019lhn})
\begin{eqnarray}
\left\langle \tilde{J}\tilde{S}\right\rangle &=&-\alpha _{0}\,,  \notag \\
\left\langle \tilde{G}_{a}\tilde{G}_{b}\right\rangle &=&\alpha _{0}\delta
_{ab}\,,  \notag \\
\left\langle \tilde{J}\tilde{M}\right\rangle &=&\,\left\langle \tilde{H}%
\tilde{S}\right\rangle =-\alpha _{1}\,,  \notag \\
\left\langle \tilde{G}_{a}\tilde{P}_{b}\right\rangle &=&\alpha _{1}\delta
_{ab}\,,  \notag \\
\left\langle \tilde{J}\tilde{T}\right\rangle &=&\,\left\langle \tilde{H}%
\tilde{M}\right\rangle =-\alpha _{2}\,.  \notag \\
\left\langle \tilde{G}_{a}\tilde{Z}_{b}\right\rangle &=&\left\langle \tilde{P%
}_{a}\tilde{P}_{b}\right\rangle =\alpha _{2}\delta _{ab}\,,  \label{MEBinvt}
\end{eqnarray}%
which can be obtained by setting $\mu _{0}=\mu _{2}=\mu _{4}=0$ and considering the following
rescaling of the relativistic parameters $\mu $'s:%
\begin{equation}
\mu _{1}=-\alpha _{0}\xi ^{2}\,,\qquad \mu _{3}=-\alpha _{1}\xi
^{2}\,,\qquad \mu _{5}=-\alpha _{2}\xi ^{2}\,.
\end{equation}
Nevertheless, at the level of the MENt algebra, the invariant tensor \eqref{MEBinvt} alone results to be degenerate. On the other hand, putting \eqref{ENinvt}, \eqref{MENtinvt}, and \eqref{MEBinvt} all together one would obtain a non-degenerate bilinear form. However, we observe that \eqref{ENinvt}, \eqref{MENtinvt}, and \eqref{MEBinvt} pertain to two different NR limits. Thus, in the sequel we will construct a CS action based on the MENt algebra by exploiting the components given by \eqref{ENinvt} and \eqref{MENtinvt} which describe non-degenerate invariant tensor for the MENt algebra. As we shall see in Section 6, the complete set of invariant tensor given by \eqref{ENinvt}, \eqref{MENtinvt}, and \eqref{MEBinvt} can alternatively be obtained through the $S$-expansion procedure allowing to write the most general CS action based on the MENt algebra.

The gauge connection one-form $A$ for the MENt algebra reads%
\begin{eqnarray}
A &=&\tau \tilde{H}+e^{a}\tilde{P}_{a}+\omega \tilde{J}+\omega ^{a}\tilde{G}%
_{a}+k\tilde{Z}+k^{a}\tilde{Z}_{a}+m\tilde{M}+s\tilde{S}+t\tilde{T}+t^{a}%
\tilde{T}_{a}+b^{a}\tilde{B}_{a}  \notag \\
&&+v^{a}\tilde{V}_{a}+b\tilde{B}+y\tilde{Y}+\varpi\tilde{W}\,. \label{connMENt}
\end{eqnarray}%
The curvature two-form $F=dA+\frac{1}{2}\left[ A,A\right] $ is given by%
\begin{eqnarray}
F &=&R\left( \tau \right) \tilde{H}+R^{a}\left( e^{b}\right) \tilde{P}%
_{a}+R\left( \omega \right) \tilde{J}+R^{a}\left( \omega ^{b}\right) \tilde{G%
}_{a}+R\left( k\right) \tilde{Z}+R^{a}\left( k^{b}\right) \tilde{Z}_{a}
\notag \\
&&+R\left( m\right) \tilde{M}+R\left( s\right) \tilde{S}+R\left( t\right)
\tilde{T}+R^{a}\left( t^{b}\right) \tilde{T}_{a}+R^{a}\left( b^{b}\right)
\tilde{B}_{a}+R^{a}\left( v^{b}\right) \tilde{V}_{a}  \notag \\
&&+R\left( b\right) \tilde{B}+R\left( y\right) \tilde{Y}+R\left( \varpi\right)
\tilde{W}\,\,, \label{curvMENt}
\end{eqnarray}%
where the explicit definition of every curvature can be found in Appendix A.

A CS gravity action based on the MENt algebra can
be constructed by combining \eqref{ENinvt} and \eqref{MENtinvt} with the gauge connection 1-form \eqref{connMENt}, and it reads, up to boundary terms, as follows:
\begin{eqnarray}
I_{\text{MENt}} &=&\frac{k}{4\pi }%
\int \Bigg \lbrace \beta _{0} \left[b_{a}R^{a}\left( \omega ^{b}\right) +\omega
_{a}R^{a}\left( b^{b}\right) -2 b R\left( \omega \right) -sds \right] \notag \\
&&+ 2 \beta _{1} \left[e_{a}R^{a}\left( b^{b}\right) + t_{a}R^{a}\left( \omega
^{b}\right) - y R\left( \omega \right) - m R\left( s\right) - \tau R\left(b\right) \right] \notag \\
&&+ \beta_2 \left[  e_{a}R^{a}\left( t^{b}\right) + t_{a}R^{a}\left( e^{b}\right) +  k_{a}R^{a}\left( b^{b}\right) + b_{a}R^{a}\left( k^{b}\right) + v_{a}R^{a}\left( \omega^{b}\right) \right.  \notag \\
&&+ \left.  \omega_{a}R^{a}\left( v^{b}\right) - 2 b R\left( k \right) - 2 \varpi R\left( \omega \right) - 2 y R\left( \tau \right) -2 tds - m dm  \right] \Bigg \rbrace \,.  \label{MENtCS}
\end{eqnarray}
The CS gravity action \eqref{MENtCS} is invariant under the MENt algebra given by \eqref{MEB} and \eqref{MENt} by construction, and it is split into three different independent terms. In particular, the term proportional to $\beta_0$ corresponds to an exotic sector of the extended Newtonian gravity term introduced in \cite{Concha:2019dqs} which can be obtained as a NR limit of an $U(1)$-extension of the exotic Einstein term \cite{Witten:1988hc}. The term proportional to $\beta_1$ is the standard extended Newtonian gravity action presented in \cite{Ozdemir:2019orp}. On the other hand, the piece proportional to $\beta_2$ contains some terms appearing in the CS action invariant under the exotic Newtonian algebra\footnote{Also known as enhanced Bargmann-Newton-Hooke algebra \cite{Bergshoeff:2020fiz}.} introduced in \cite{Concha:2019dqs}, together with some completely new terms. 

It is interesting to note that an enhancement of the Maxwell algebra allows to construct a Maxwellian version of the extended Newtonian gravity containing known NR gravity theories as sub-cases. Although the field content is bigger than the extended Newtonian one, the present NR theory does not contain a cosmological constant term. As we shall see in the next section, in order to accommodate a cosmological constant to \eqref{MENtCS}, it is necessary to consider a different NR algebra which can be obtained as an IW contraction of the relativistic [enhanced AdS-$\mathcal{L}$] $\oplus \mathfrak{u}(1)^{3}$ algebra. The NR limit and a possible vanishing cosmological constant limit are not the only ways to obtain the present MENt gravity theory. Indeed, as we shall show, the MENt algebra and its most general invariant tensor can alternatively be found using the semigroup expansion method.

The non-degeneracy of the invariant tensor has allowed to achieve a well-defined NR CS gravity action whose equations of motion are given by the vanishing of all the curvatures \eqref{curvMENt}.

The CS action \eqref{MENtCS} can equivalently be obtained as NR limit of \eqref{EMaxCS}. To see this, one shall express the relativistic enhanced Maxwell 1-form fields as a linear combination of the NR ones as
\begin{eqnarray}
W^{0} &=& \omega + \frac{1}{2\xi^2} s - \frac{1}{2\xi^4} b \,,\qquad \quad W^{a}= \frac{1}{\xi} \omega^a - \frac{1}{\xi^3} b^a \,,  \notag \\
E^{0} &=& \tau + \frac{1}{2\xi^2} m - \frac{1}{2\xi^4} y \,,\qquad \quad E^{a}= \frac{1}{\xi} e^a - \frac{1}{\xi^3} t^a \,,  \notag \\
K^{0} &=& k + \frac{1}{2\xi^2} t - \frac{1}{2\xi^4} \varpi \,,\qquad \quad K^{a}= \frac{1}{\xi} k^a - \frac{1}{\xi^3} v^a \,,  \notag \\
\Sigma^{0} &=& - \frac{1}{\xi^2} s \,,\qquad \qquad \qquad \quad \quad \ \Sigma^{a}= - \frac{1}{2\xi} \omega^a - \frac{1}{2\xi^3} b^a \,, \notag  \\
L^{0} &=& - \frac{1}{\xi^2} m \,,\qquad \qquad \qquad \quad \quad L^{a}= - \frac{1}{2\xi} e^a - \frac{1}{2\xi^3} t^a \,,  \notag \\
\Gamma^{0} &=& - \frac{1}{\xi^2} t \,,\qquad \qquad \qquad \quad \quad \ \, \Gamma^{a}= - \frac{1}{2\xi} k^a - \frac{1}{2\xi^3} v^a \,.  \label{NR1fields}
\end{eqnarray}
and the gauge fields dual to the $\mathfrak{u}(1)$ generators $Y_1$, $Y_2$, and $Y_3$ (namely $S$, $M$, and $T$, respectively) as
\begin{equation}
S= \omega - \frac{1}{2\xi^2} s + \frac{1}{2\xi^4} b \,,\qquad M= \tau - \frac{1}{2\xi^2} m + \frac{1}{2\xi^4} y \,,\qquad T= k - \frac{1}{2\xi^2} t + \frac{1}{2\xi^4} \varpi \,.
\label{NR2fields}
\end{equation}
Then, substituting back \eqref{NR1fields} and \eqref{NR2fields}, together with \eqref{betasMENt}, in \eqref{EMaxCS}, omitting boundary terms and taking the limit $\xi \rightarrow \infty$, one precisely recovers \eqref{MENtCS}.

Finally, one can see that each independent term of the action \eqref{MENtCS} is
invariant under the gauge transformation laws $\delta A = d \lambda + \left[ A,\lambda \right]$, being
\begin{eqnarray}
\lambda &=&\Lambda \tilde{H}+\Lambda^{a}\tilde{P}_{a}+\Omega \tilde{J}+\Omega ^{a}\tilde{G}%
_{a}+\kappa \tilde{Z}+\kappa^{a}\tilde{Z}_{a}+ \chi \tilde{M}+ \varsigma \tilde{S}+ \pi \tilde{T}+\pi^{a}%
\tilde{T}_{a}+\rho^{a}\tilde{B}_{a}  \notag \\
&&+\nu^{a}\tilde{V}_{a}+\rho\tilde{B}+\gamma \tilde{Y}+\varrho\tilde{W} \label{gaugeparMENt}
\end{eqnarray}
the gauge parameter. The gauge transformations of the theory can be found in Appendix B.

\section{Enlarged extended Newtonian gravity and flat limit}

In this section, we apply a NR limit to the [enhanced
AdS-$\mathcal{L}$] $\, \oplus \,
\mathfrak{u}\left( 1\right) ^3 $ algebra previously introduced. We show that the new NR algebra can be seen as an enlargement of the extended Newtonian algebra \cite{Ozdemir:2019orp} (which we have denoted as EEN algebra) and
reproduces the MENt algebra (\ref{MEB})-(\ref{MENt}) in
the vanishing cosmological constant limit $\ell \rightarrow \infty $.

\subsection{Enlarged extended Newtonian algebra and non-relativistic limit}

In the previous section we have constructed a Maxwellian version of the
extended Newtonian algebra by applying an Inönü-Wigner contraction to the
[enhanced Maxwell] $\, \oplus \,
\mathfrak{u}\left( 1\right)^3 $ algebra (\ref{EMax}). In order to obtain a
NR version of the [enhanced
AdS-$\mathcal{L}$] $\, \oplus \,
\mathfrak{u}\left( 1\right) ^3 $ algebra (where the latter is given by (\ref{EMax}) and (\ref{EAdSL})), we
consider the same redefinitions of the relativistic generators as in (\ref%
{NR1})-(\ref{NR2}) which provides us with a well-defined $\xi \rightarrow
\infty $ limit. The new NR algebra is then generated by the
set of generators of the EEB algebra $\left\{ \tilde{J%
},\tilde{G}_{a},\tilde{S},\tilde{H},\tilde{P}_{a},\tilde{M},\tilde{Z},\tilde{%
Z}_{a},\tilde{T}\right\} $, a set of additional generators $\left\{ \tilde{T}%
_{a},\tilde{B}_{a},\tilde{V}_{a}\right\} $, and three central charges given
by $\tilde{B}$, $\tilde{Y}$, $\tilde{W}$. Such generators satisfy the
commutation relations (\ref{MEB})-(\ref{MENt}) along with%
\begin{eqnarray}
\left[ \tilde{H},\tilde{Z}_{a}\right] &=&\frac{1}{\ell ^{2}}\epsilon _{ab}%
\tilde{P}_{b}\,,\,\left[ \tilde{P}_{a},\tilde{Z}_{b}\right] =-\frac{1}{\ell
^{2}}\epsilon _{ab}\tilde{M}\,,\,\left[ \tilde{Z},\tilde{P}_{a}\right] =%
\frac{1}{\ell ^{2}}\epsilon _{ab}\tilde{P}_{b}\,,  \notag \\
\left[ \tilde{Z},\tilde{Z}_{a}\right] &=&\frac{1}{\ell ^{2}}\epsilon _{ab}%
\tilde{Z}_{b}\,,\,\left[ \tilde{Z}_{a},\tilde{Z}_{b}\right] =-\frac{1}{\ell
^{2}}\epsilon _{ab}\tilde{T}\,,\,  \label{EEB}
\end{eqnarray}%
and%
\begin{eqnarray}
\left[ \tilde{H},\tilde{V}_{a}\right] &=&\frac{1}{\ell ^{2}}\epsilon _{ab}%
\tilde{T}_{b}\,,\,\left[ \tilde{P}_{a},\tilde{V}_{b}\right] =-\frac{1}{\ell
^{2}}\epsilon _{ab}\tilde{Y}\,,\ \, \left[ \tilde{Z},\tilde{T}_{a}\right] =%
\frac{1}{\ell ^{2}}\epsilon _{ab}\tilde{T}_{b}\,,  \notag \\
\,\,\left[ \tilde{Z},\tilde{V}_{a}\right] &=&\frac{1}{\ell ^{2}}\epsilon
_{ab}\tilde{V}_{b}\,,\,\left[ \tilde{Z}_{a},\tilde{T}_{b}\right] =-\frac{1}{%
\ell ^{2}}\epsilon _{ab}\tilde{Y}\,,\,\left[ \tilde{M},\tilde{Z}_{a}\right] =%
\frac{1}{\ell ^{2}}\epsilon _{ab}\tilde{T}_{b}\,,  \notag \\
\left[ \tilde{T},\tilde{P}_{a}\right] &=&\frac{1}{\ell ^{2}}\epsilon _{ab}%
\tilde{T}_{b}\,,\,\left[ \tilde{Z}_{a},\tilde{V}_{b}\right] =-\frac{1}{\ell
^{2}}\epsilon _{ab}\tilde{W}\,,\,\left[ \tilde{T},\tilde{Z}_{a}\right] =%
\frac{1}{\ell ^{2}}\epsilon _{ab}\tilde{V}_{b}\,.  \label{EEN}
\end{eqnarray}%
Note that the commutation relations (\ref{MEB}) and (%
\ref{EEB}) define a subalgebra corresponding to the EEB algebra introduced in \cite{Concha:2019lhn}. The novel non-relativistic algebra obtained here can be seen as an
enlargement of the extended Newtonian algebra, and we will denote it as EEN
algebra. An interesting feature of such NR algebra is given by
the explicit presence of a scale $\ell $ which allows us to accommodate a
cosmological constant into the Maxwellian version of the extended Newtonian
gravity (\ref{MEB})-(\ref{MENt}). Naturally, the vanishing cosmological constant limit $\ell
\rightarrow \infty $ reproduces the MENt algebra. 

Let us note that the EEN algebra can be rewritten as
three copies of the enhanced Nappi-Witten algebra defined in \cite{Bergshoeff:2020fiz}.
Indeed, the EEN algebra can be written as%
\begin{eqnarray}
\left[ J^{\pm },G_{a}^{\pm }\right]  &=&\epsilon _{ab}G_{b}^{\pm }\,,\quad \ %
\left[ G_{a}^{\pm },G_{b}^{\pm }\right] =-\epsilon _{ab}S^{\pm }\,,\quad %
\left[ S^{\pm },G_{a}^{\pm }\right] =\epsilon _{ab}B_{b}^{\pm }\,,  \notag \\
\left[ J^{\pm },B_{a}^{\pm }\right]  &=&\epsilon _{ab}B_{b}^{\pm }\,,\quad \ %
\left[ G_{a}^{\pm },B_{b}^{\pm }\right] =-\epsilon _{ab}B^{\pm }\,,  \notag
\\
\left[ \hat{J},\hat{G}_{a}\right]  &=&\epsilon _{ab}\hat{G}_{b}\,,\quad \ %
\left[ \hat{G}_{a},\hat{G}_{b}\right] =-\epsilon _{ab}\hat{S}\,,\quad \ \ %
\left[ \hat{S},\hat{G}_{a}\right] =\epsilon _{ab}\hat{B}_{b}\,, \notag \\
\left[ \hat{J},\hat{B}_{a}\right]  &=&\epsilon _{ab}\hat{B}_{b}\,,\quad \ %
\left[ \hat{G}_{a},\hat{B}_{b}\right] =-\epsilon _{ab}\hat{B}\,,
\end{eqnarray}%
by considering the following redefinition of the generators:
\begin{equation*}
\begin{tabular}{ccc}
$\tilde{J}=\hat{J}+J^{+}+J^{-}\,,$ & $\tilde{H}=1/\ell \,\left(
J^{+}-J^{-}\right) \,,$ & $\tilde{Z}=1/\ell ^{2}\,\left( J^{+}+J^{-}\right)
\,,$ \\
$\tilde{G}_{a}=\hat{G}_{a}+G_{a}^{+}+G_{a}^{-}\,,$ & $\tilde{P}_{a}=1/\ell
\,\left( G_{a}^{+}-G_{a}^{-}\right) \,,$ & $\tilde{Z}_{a}=1/\ell
^{2}\,\left( G_{a}^{+}+G_{a}^{-}\right) \,,$ \\
$\tilde{S}=\hat{S}+S^{+}+S^{-}\,,$ & $\tilde{M}=1/\ell \,\left(
S^{+}-S^{-}\right) \,,$ & $\tilde{T}=1/\ell ^{2}\,\left( S^{+}+S^{-}\right)
\,,$ \\
$\tilde{B}_{a}=\hat{B}_{a}+B_{a}^{+}+B_{a}^{-}\,,$ & $\ \tilde{T}_{a}=1/\ell
\,\left( B_{a}^{+}-B_{a}^{-}\right) \,,$ & $\tilde{V}_{a}=1/\ell
^{2}\,\left( B_{a}^{+}+B_{a}^{-}\right) \,,$ \\
$\tilde{B}=\hat{B}+B^{+}+B^{-}\,,\,$ & $\tilde{Y}=1/\ell \,\left(
B^{+}-B^{-}\right) $ & $\tilde{W}=1/\ell ^{2}\,\left( B^{+}+B^{-}\right) \,.$%
\end{tabular}%
\end{equation*}%
On the other hand, a different redefinition of the generators of the EEN
algebra can be considered. In particular, the redefinition%
\begin{equation*}
\begin{tabular}{lll}
$\tilde{J}=\mathcal{J}+\hat{J}\,,$ & $\tilde{H}=\mathcal{H}\,,$ & $\tilde{Z}%
=1/\ell ^{2}\,\mathcal{J}\,,$ \\
$\tilde{G}_{a}=\mathcal{G}_{a}+\hat{G}_{a}\,,\,$ & $\tilde{P}_{a}=\mathcal{P}%
_{a}\,,$ & $\tilde{Z}_{a}=1/\ell ^{2}\,\mathcal{G}_{a}\,,$ \\
$\tilde{S}=\mathcal{S}+\hat{S}\,,$ & $\tilde{M}=\mathcal{M}\,,$ & $\tilde{T}%
=1/\ell ^{2}\,\mathcal{S}\,,$ \\
$\tilde{B}_{a}=\mathcal{B}_{a}+\hat{B}_{a}\,,$ & $\tilde{T}_{a}=\mathcal{T}%
_{a}\,,$ & $\tilde{V}_{a}=1/\ell ^{2}\,\mathcal{B}_{a}\,,$ \\
$\tilde{B}=\mathcal{B}+\hat{B}\,,$ & $\tilde{Y}=\mathcal{Y}\,,$ & $\tilde{W}%
=1/\ell ^{2}\mathcal{B}\,,$%
\end{tabular}%
\end{equation*}%
allows to rewrite the EEN algebra (\ref{MEB}), (\ref{MENt}), (%
\ref{EEB}), and (\ref{EEN}) as the direct sum of the exotic Newtonian algebra \cite{Concha:2019dqs},%
\begin{eqnarray}
\left[ \mathcal{J},\mathcal{G}_{a}\right]  &=&\epsilon _{ab}\mathcal{G}%
_{b}\,,\ \left[ \mathcal{G}_{a},\mathcal{G}_{b}\right] =-\epsilon _{ab}%
\mathcal{S}\,,\ \ \left[ \mathcal{H},\mathcal{G}_{a}\right] =\epsilon _{ab}%
\mathcal{P}_{b}\,,  \notag \\
\left[ \mathcal{J},\mathcal{P}_{a}\right]  &=&\epsilon _{ab}\mathcal{P}%
_{b}\,,\,\left[ \mathcal{G}_{a},\mathcal{P}_{b}\right] =-\epsilon _{ab}%
\mathcal{M}\,,\,\left[ \mathcal{H},\mathcal{B}_{a}\right] =\epsilon _{ab}%
\mathcal{T}_{b}\,,  \notag \\
\left[ \mathcal{J},\mathcal{B}_{a}\right]  &=&\epsilon _{ab}\mathcal{B}%
_{b}\,,\,\left[ \mathcal{G}_{a},\mathcal{B}_{b}\right] =-\epsilon _{ab}%
\mathcal{B}\,,\ \ \, \left[ \mathcal{J},\mathcal{T}_{a}\right] =\epsilon _{ab}%
\mathcal{T}_{b}\,,  \notag \\
\left[ \mathcal{S},\mathcal{G}_{a}\right]  &=&\epsilon _{ab}\mathcal{B}%
_{b}\,,\,\,\left[ \mathcal{G}_{a},\mathcal{T}_{b}\right] =\epsilon _{ab}%
\mathcal{Y}\,,\quad \ \, \left[ \mathcal{S},\mathcal{P}_{a}\right] =\epsilon _{ab}%
\mathcal{T}_{b}\,, \notag  \\
\left[ \mathcal{M},\mathcal{G}_{a}\right]  &=&\epsilon _{ab}\mathcal{T}%
_{b}\,,\,\left[ \mathcal{P}_{a},\mathcal{B}_{b}\right] =\epsilon _{ab}%
\mathcal{Y}\,,\quad \ \left[ \mathcal{H},\mathcal{P}_{a}\right] =\frac{1}{\ell ^{2}%
}\epsilon _{ab}\mathcal{G}_{b}\,,  \notag \\
\left[ \mathcal{H},\mathcal{T}_{a}\right]  &=&\frac{1}{\ell ^{2}}\epsilon
_{ab}\mathcal{B}_{b}\,,\,\left[ \mathcal{P}_{a},\mathcal{P}_{b}\right] =-%
\frac{1}{\ell ^{2}}\epsilon _{ab}\mathcal{S}\,,  \notag \\
\left[ \mathcal{M},\mathcal{P}_{a}\right]  &=&\frac{1}{\ell ^{2}}\epsilon
_{ab}\mathcal{B}_{b}\,,\,\left[ \mathcal{P}_{a},\mathcal{T}_{b}\right] =-%
\frac{1}{\ell ^{2}}\epsilon _{ab}\mathcal{B}\,,  \label{eN}
\end{eqnarray}%
and the enhanced Nappi-Witten algebra,
\begin{eqnarray}
\left[ \hat{J},\hat{G}_{a}\right]  &=&\epsilon _{ab}\hat{G}_{b}\,,\quad \ %
\left[ \hat{G}_{a},\hat{G}_{b}\right] =-\epsilon _{ab}\hat{S}\,,\quad \ \ %
\left[ \hat{S},\hat{G}_{a}\right] =\epsilon _{ab}\hat{B}_{b}\,,  \notag \\
\left[ \hat{J},\hat{B}_{a}\right]  &=&\epsilon _{ab}\hat{B}_{b}\,,\quad \ %
\left[ \hat{G}_{a},\hat{B}_{b}\right] =-\epsilon _{ab}\hat{B}\,.  \label{ENW}
\end{eqnarray}%
The exotic Newtonian algebra has been recently introduced in \cite{Concha:2019dqs} and
subsequently studied in \cite{Bergshoeff:2020fiz} and allows us to accommodate a cosmological
constant into the extended Newtonian gravity theory. One can see that, as we
have shown previously, the same behavior appears at the relativistic level.
Indeed the enhanced AdS-Lorentz algebra can also be rewritten as
three copies of the Poincaré algebra and as the direct sum of the coadjoint
AdS and Poincaré algebra after an appropriate redefinition of the
generators. In particular, the enhanced Nappi-Witten algebra (\ref{ENW})
appears as an IW contraction of the Poincaré $\oplus \, \mathfrak{u}%
\left( 1\right) $ algebra, while the exotic Newtonian algebra can be
obtained as a NR limit of the coadjoint AdS algebra \cite{Bergshoeff:2020fiz}.

In what follows we present the NR CS action based on the
EEN algebra (\ref{MEB}), (\ref{MENt}), (\ref{EEB}), and (\ref{EEN}). Such basis is preferred in order to make clear the flat
limit leading to the MENt gravity. Furthermore, as
we shall see, such basis present an alternative way to include a
cosmological constant into a three-dimensional NR CS gravity
action diverse to the one discussed in \cite{Concha:2019dqs}.

\subsection{Non-relativistic EEN Chern-Simons gravity action and flat limit}

A well-defined CS action requires invariant non-degenerate bilinear form.
Interestingly, the presence of the central charges $\tilde{B}$, $\tilde{Y}$, and $\tilde{W}$ assures to have a non-degenerate invariant tensor for the EEN
algebra. In particular, the non-vanishing components of the invariant tensor
for the EEN algebra are given by the MENt ones (\ref{ENinvt})-(\ref{MENtinvt})
along with%
\begin{eqnarray}
\left\langle \tilde{M}\tilde{T}\right\rangle &=&\left\langle \tilde{Z}\tilde{%
Y}\right\rangle =\left\langle \tilde{H}\tilde{W}\right\rangle =-\frac{\beta
_{1}}{\ell ^{2}}\,,  \notag \\
\left\langle \tilde{P}_{a}\tilde{V}_{b}\right\rangle &=&\left\langle \tilde{Z%
}_{a}\tilde{T}_{b}\right\rangle =\frac{\beta _{1}}{\ell ^{2}}\delta _{ab}\,,
\notag \\
\left\langle \tilde{T}\tilde{T}\right\rangle &=&\left\langle \tilde{Z}\tilde{%
W}\right\rangle =-\frac{\beta _{2}}{\ell ^{2}}\,, \notag  \\
\left\langle \tilde{Z}_{a}\tilde{V}_{b}\right\rangle &=&\frac{\beta _{2}}{%
\ell ^{2}}\delta _{ab}\,,  \label{EENinvt}
\end{eqnarray}%
where the relativistic parameters $\mu $'s have been rescaled as
\begin{equation}\label{betasEEN}
\mu _{0}=\mu _{1}=-\beta _{0}\xi ^{4}\,,\qquad \mu _{2}=\mu _{3}=-\beta
_{1}\xi ^{4}\,,\qquad \mu _{4}=\mu _{5}=-\beta _{2}\xi ^{4}\, .
\end{equation}%
On the other hand, one can show that the EEN algebra
can also admit the MEB invariant tensor (\ref{MEBinvt}) along with%
\begin{eqnarray}
\left\langle \tilde{H}\tilde{T}\right\rangle &=&\left\langle \tilde{Z}\tilde{%
M}\right\rangle =-\frac{\alpha _{1}}{\ell ^{2}}\,,  \notag \\
\left\langle \tilde{P}_{a}\tilde{Z}_{b}\right\rangle &=&\frac{\alpha _{1}}{%
\ell ^{2}}\delta _{ab}\,,  \notag \\
\left\langle \tilde{Z}\tilde{T}\right\rangle &=&-\frac{\alpha _{2}}{\ell ^{2}%
}\,, \notag  \\
\left\langle \tilde{Z}_{a}\tilde{Z}_{b}\right\rangle &=&\frac{\alpha _{2}}{%
\ell ^{2}}\delta _{ab}\,,  \label{EEBinvt}
\end{eqnarray}%
when we set $\mu _{0}=\mu _{2}=\mu _{4}=0$ and consider the following
rescaling of the relativistic parameters $\mu $'s:%
\begin{equation}
\mu _{1}=-\alpha _{0}\xi ^{2}\,,\qquad \mu _{3}=-\alpha _{1}\xi
^{2}\,,\qquad \mu _{5}=-\alpha _{2}\xi ^{2}\,.
\end{equation}%
One can see that the components of the invariant tensor proportional to the $\alpha $'s are the respective components of the invariant tensor of the EEB
algebra introduced in \cite{Concha:2019lhn}. On the other hand, those related to the $\beta $'s shall reproduce the enlarged extended Newtonian gravity action. As it was discussed in the MENt case, the NR limit does not allow to have both types of invariant tensor. In this section we shall focus on those proportional to the $\beta$'s constants allowing to construct a novel NR CS action. In Section 6, we shall see that the complete set containing both the $\alpha$'s and the $\beta$'s constants appearing in the full invariant tensor can be properly obtained considering the $S$-expansion method.

Although the gauge connection one-form $A$ is the same than the MENt one,
the corresponding curvature two-form $F$ is different due to the presence of
new commutation relations involving the $\ell$ parameter. Indeed, the curvature two-form $F$ reads
\begin{eqnarray}
F &=&R\left( \tau \right) \tilde{H}+\hat{R}^{a}\left( e^{b}\right) \tilde{P}%
_{a}+R\left( \omega \right) \tilde{J}+R^{a}\left( \omega ^{b}\right) \tilde{G%
}_{a}+R\left( k\right) \tilde{Z}+\hat{R}^{a}\left( k^{b}\right) \tilde{Z}_{a}
\notag \\
&&+\hat{R}\left( m\right) \tilde{M}+R\left( s\right) \tilde{S}+\hat{R}\left(
t\right) \tilde{T}+\hat{R}^{a}\left( t^{b}\right) \tilde{T}_{a}+R^{a}\left(
b^{b}\right) \tilde{B}_{a}+\hat{R}^{a}\left( v^{b}\right) \tilde{V}_{a}
\notag \\
&&+R\left( b\right) \tilde{B}+\hat{R}\left( y\right) \tilde{Y}+\hat{R}\left(
\varpi\right) \tilde{W}\,\,,
\end{eqnarray}%
where the explicit expression of every curvature two-form is given in Appendix A.

A CS gravity action based on the EEN algebra can
be constructed by combining \eqref{ENinvt}-\eqref{MENtinvt} and \eqref{EENinvt} with the gauge connection 1-form \eqref{connMENt}, and it reads, up to boundary terms, as follows:
\begin{eqnarray}
I_{\text{EEN}} &=& \frac{k}{4\pi }%
\int \Bigg \lbrace \beta _{0} \left[ b_{a}R^{a}\left( \omega ^{b}\right) +\omega
_{a}R^{a}\left( b^{b}\right) -2 b R\left( \omega \right) -sds \right] \notag \\
&&+  \beta _{1} \Bigg [  e_{a}R^{a}\left( b^{b}\right) + b_{a}\hat{R}^{a}\left( e^{b}\right) +  t_{a}R^{a}\left( \omega
^{b}\right) + \omega_{a}\hat{R}^{a}\left(t^{b}\right) - 2 y R\left( \omega \right) - 2 b R\left( \tau \right) - 2 m ds  \notag \\
&&+  \frac{1}{\ell^{2}}\left( t_{a}\hat{R}^{a}\left(k^{b}\right) +  v_{a}\hat{R}^{a}\left( e^{b}\right) + e_{a}\hat{R}^{a}\left(v^{b}\right)+k_{a}\hat{R}^{a}\left(t^{b}\right)
-2yR\left(k\right)-2\varpi R\left(\tau\right) - 2m dt \right) \Bigg ] \notag \\
&&+ \beta_2 \Bigg [  e_{a}\hat{R}^{a}\left( t^{b}\right) + t_{a}\hat{R}^{a}\left( e^{b}\right) +  k_{a}R^{a}\left( b^{b}\right) + b_{a}\hat{R}^{a}\left( k^{b}\right) + v_{a}R^{a}\left( \omega^{b}\right)   \notag \\
&&+  \omega_{a}\hat{R}^{a}\left( v^{b}\right) - 2 b R\left( k \right) - 2 \varpi R\left( \omega \right) - 2 y R\left( \tau \right) -2 tds - m dm    \notag \\
&&+ \frac{1}{\ell^{2}} \left( v_{a}\hat{R}^{a}\left(k^{b}\right) + k_{a}\hat{R}^{a}\left(v^{b}\right) - 2 \varpi R\left(k\right) - t dt  \right) \Bigg ] \Bigg \rbrace \,. \label{EENCS}
\end{eqnarray}
The CS action \eqref{EENCS} is invariant under the EEN algebra by construction.
One can notice, looking also at \eqref{MENtCS}, that there are three independent sectors proportional to the different coupling constants $\beta_0$, $\beta_1$, and $\beta_2$. The term proportional to $\beta_0$ is exactly the same as in \eqref{MENtCS}. On the other hand, the terms proportional to $\beta_1$ and $\beta_2$ involve, besides the MENt contributions, also new pieces. In the flat limit $\ell \rightarrow \infty$ of \eqref{EENCS} we recover the MENt gravity action \eqref{MENtCS}. Observe that our construction gives an alternative way to include a cosmological constant into the three-dimensional NR CS gravity action \eqref{MENtCS}.

The non-degeneracy of the invariant tensor has allowed to achieve a well-defined CS gravity action whose equations of motion are given by the vanishing of all the curvatures \eqref{curvEEN}.

The CS action \eqref{EENCS} we have just constructed can equivalently be obtained as NR limit of the CS action \eqref{EAdSLCS} based on the [enhanced AdS-$\mathcal{L}$] $\oplus\, \mathfrak{u}(1)^{3}$ algebra. Indeed, expressing the relativistic enhanced AdS-$\mathcal{L}$ 1-form fields in \eqref{EAdSLCS} as a linear combination of the NR ones as in \eqref{NR1fields} and the gauge fields $S$, $M$, and $T$ as in \eqref{NR2fields}, together with \eqref{betasEEN}, and taking the limit $\xi \rightarrow \infty$, one precisely recovers the NR CS action \eqref{EENCS} for the EEN algebra.

Concluding, we can see that each independent term of the action \eqref{EENCS} is
invariant under the gauge transformation laws $\delta A = d \lambda + \left[ A,\lambda \right]$ with gauge parameter $\lambda$ given by \eqref{gaugeparMENt}. Specifically, the explicit gauge transformations of the theory are defined in Appendix B.

\section{Non-relativistic algebras and semigroup expansion method}

In this section, we present an alternative procedure to recover the new
NR algebras previously introduced. In particular, we show that
the MENt and EEN algebras can be alternatively obtained by considering the $S$-expansion procedure. The Lie algebra expansion method has first been introduced in \cite{Hatsuda:2001pp} in the context of AdS superstring and then developed by considering the Maurer-Cartan equations in \cite{deAzcarraga:2002xi,deAzcarraga:2004zj,deAzcarraga:2007et}. The expansion based on semigroups has then been introduced in \cite{Izaurieta:2006zz} and subsequently developed in \cite{Caroca:2011qs,Andrianopoli:2013ooa,Artebani:2016gwh,Ipinza:2016bfc,Inostroza:2017ezc,Inostroza:2018gzd}. The $S$-expansion consists in defining a new expanded Lie algebra $\mathfrak{G}=S\times\mathfrak{g}$, by combining the elements of the structure constants of a Lie algebra $\mathfrak{g}$ with the semigroup $S$. Such procedure provides us not only with the commutation relations, but also with the complete set of non-vanishing components of the invariant tensor for the expanded algebra. At the NR level, the Lie algebra expansion method has been used by diverse authors leading to interesting results \cite{Bergshoeff:2019ctr,deAzcarraga:2019mdn,Romano:2019ulw,Gomis:2019nih,Kasikci:2020qsj,Fontanella:2020eje}.

Here we shall consider the $S$-expansion of the enhanced Nappi-Witen algebra \eqref{ENW} formerly discussed. The motivation to consider the enhanced Nappi-Witten algebra as the original algebra is twofold. First, as was shown in \cite{Concha:2019lhn,Penafiel:2019czp}, the $S$-expansion of the Nappi-Witten algebra allows us to recover diverse NR algebras. Second, since the  $\mathfrak{iso}(2,1) \oplus \,\mathfrak{u}(1)$ algebra can be used to obtain the respective relativistic counterparts, it seems natural to expect the same behavior for the NR version of the $\mathfrak{iso}(2,1) \oplus \,\mathfrak{u}(1)$ algebra, which is given by the enhanced Nappi-Witten algebra. 

Let us first recall the enhanced Nappi-Witten algebra $\mathfrak{g}$, whose generators satisfy
\begin{eqnarray}
\left[ J,G_{a}\right] &=&\epsilon _{ab}G_{b}\,,\quad \left[ G_{a},G_{b}%
\right] =-\epsilon _{ab}S\,,\quad \ \ \left[ S,G_{a}\right] =\epsilon
_{ab}B_{b}\,,  \notag \\
\left[ J,B_{a}\right] &=&\epsilon _{ab}B_{b}\,,\quad \ \left[ G_{a},B_{b}%
\right] =-\epsilon _{ab}B\,.  \label{eNW}
\end{eqnarray}%
This algebra appears as a particular IW contraction of the $\mathfrak{iso}(2,1) \oplus \,\mathfrak{u}(1)$ algebra. Indeed, as was mentioned in \cite{Bergshoeff:2020fiz}, one can write
the relativistic Poincaré and $\mathfrak{u}\left( 1\right) $ generators $%
\left\{ J_{A},P_{A},Y_{1}\right\} $ in terms of the enhanced Nappi-Witten
ones as%
\begin{eqnarray}
J_{0} &=&\frac{J}{2}-\xi ^{4}B\,,\qquad \quad \ \ \ \ J_{a}=\frac{\xi }{2}%
G_{a}-\frac{\xi ^{3}}{2}B_{a}\,,  \notag \\
P_{0} &=&-\xi ^{2}S-\xi ^{4}B\,,\qquad \quad P_{a}=-\xi G_{a}-\xi
^{3}B_{a}\,, \notag \\
Y_{1} &=&\frac{J}{2}+\xi ^{4}B\,.
\end{eqnarray}%
Then, the aforesaid algebra (\ref{eNW}) is revealed after applying
the limit $\xi \rightarrow \infty $. Naturally, the Nappi-Witten algebra \cite{Nappi:1993ie,Figueroa-OFarrill:1999cmq} spanned by $\{J,G_{a},S \}$ appears as a subalgebra.

Let us now consider $S_{E}^{\left( 2\right) }=\left\{ \lambda _{0},\lambda
_{1},\lambda _{2},\lambda _{3}\right\} $ as the relevant semigroup whose
elements satisfy the following multiplication law,%
\begin{equation}
\lambda _{\alpha }\lambda _{\beta }=\left\{
\begin{array}{lcl}
\lambda _{\alpha +\beta }\,\,\,\, & \mathrm{if}\,\,\,\,\alpha +\beta \leq 3 \, ,
&  \\
\lambda _{3}\,\,\, & \mathrm{if}\,\,\,\,\alpha +\beta >3 \, , &
\end{array}%
\right.  \label{ml3}
\end{equation}%
where $\lambda _{3}=0_{s}$ is the zero element of the semigroup such that $%
0_{s}\lambda _{\alpha }=0_{s}$. Then, after considering a $0_{s}$-reduction of the expanded algebra $%
S_{E}^{\left( 2\right) } \times \mathfrak{g} $ ($\mathfrak{g}$ being the enhanced Nappi-Witten algebra), we find an expanded algebra spanned by the generators $\left\{ \tilde{J},\tilde{G}_{a},\tilde{S%
},\tilde{H},\tilde{P}_{a},\tilde{M},\tilde{Z},\tilde{Z}_{a},\tilde{T}%
\right\} $ and $\left\{ \tilde{T}_{a},\tilde{B}_{a},\tilde{V}_{a},\tilde{B},%
\tilde{Y},\tilde{W}\right\} $ which are related to the enhanced Nappi-Witten
ones through the semigroup elements as%
\begin{eqnarray}
\tilde{J}&=&\lambda _{0}J\,, \qquad \ \ \tilde{H}=\lambda _{1}J\,,\qquad \ \  \tilde{Z}%
=\lambda _{2}J\,, \notag\\
\tilde{G}_{a}&=&\lambda _{0}G_{a}\,,\qquad \tilde{P}_{a}=\lambda _{1}G_{a}\,, \qquad
 \tilde{Z}_{a}=\lambda _{2}G_{a}\,, \notag\\
\tilde{S}&=&\lambda _{0}S\,,\qquad \ \ \tilde{M}=\lambda _{1}S\,, \qquad \ \ \tilde{T}%
=\lambda _{2}S\,,  \notag\\
\tilde{B}_{a}&=&\lambda _{0}B_{a}\,, \qquad \tilde{T}_{a}=\lambda _{1}B_{a}\,, \qquad \tilde{V}_{a}=\lambda _{2}B_{a}\,, \notag\\
\tilde{B}&=&\lambda _{0}B\,, \qquad \ \ \tilde{Y}=\lambda _{1}B\,, \qquad \ \ \tilde{W}%
=\lambda _{2}B\,. \label{def}%
\end{eqnarray}
Using the multiplication law of the semigroup (\ref{ml3}) and the enhanced
Nappi-Witten commutators (\ref{eNW}), one finds that the expanded generators
satisfy the MENt algebra (\ref{MEB}) and (\ref{MENt}). Additionally, one can express the non-vanishing components of the
invariant tensor of the $S$-expanded algebra in terms of the original ones.
Interestingly, considering the Theorem VII. of \cite{Izaurieta:2006zz}, one can show that the
non-vanishing components of the invariant tensor for the MENt algebra are given not only by (\ref{ENinvt})-(\ref{MENtinvt}) but
also by the MEB ones (\ref{MEBinvt}). This can be seen directly from the
invariant tensor of the enhanced Nappi-Witten algebra whose components are given by%
\begin{eqnarray}
\left\langle JS\right\rangle &=&-\gamma _{1}\,,  \notag \\
\left\langle G_{a}G_{b}\right\rangle &=&\gamma _{1}\delta _{ab}\,,  \notag \\
\left\langle SS\right\rangle &=&\left\langle SB\right\rangle =-\gamma _{2}\,, \notag
\\
\left\langle G_{a}B_{b}\right\rangle &=&\gamma _{2}\delta _{ab}\,. \label{invtENW}
\end{eqnarray}%
The expanded invariant tensor coming from $\left\langle JS\right\rangle $
and $\left\langle G_{a}G_{b}\right\rangle $ generate the non-vanishing
components of the invariant tensor for the MEB algebra proportional to $%
\alpha $'s given by (\ref{MEBinvt}). In particular, the $\alpha $'s constants can be expressed in terms of $\gamma _{1}$ as%
\begin{equation}
\alpha _{0}=\lambda _{0}\gamma _{1}\,,\qquad \alpha _{1}=\lambda _{1}\gamma
_{1}\,,\qquad \alpha _{2}=\lambda _{2}\gamma _{1}\,.  \label{alpha}
\end{equation}
The invariant tensor for the MENt algebra (\ref{ENinvt})-(\ref{MENtinvt}),
proportional to the $\beta $'s, are obtained from $\left\langle
SS\right\rangle $, $\left\langle SB\right\rangle $, and $\left\langle
G_{a}B_{b}\right\rangle $ with
\begin{equation}
\beta _{0}=\lambda _{0}\gamma _{2}\,,\qquad \beta _{1}=\lambda _{1}\gamma
_{2}\,,\qquad \beta _{2}=\lambda _{2}\gamma _{2}\,.  \label{beta}
\end{equation}%
As we have previously discussed, both classes of invariant tensor can be obtained separately by considering different NR limits. Although the complete set of invariant tensor (\ref{ENinvt})-(\ref{MENtinvt}) along with (\ref{MEBinvt}) are non-degenerate for the MENt algebra, they cannot be obtained simultaneously considering a unique NR limit. Here, the semigroup expansion procedure provides us with both components of the invariant tensor. In particular, the MEB ones are related to the $\gamma_1$ appearing in the invariant tensor of the enhanced Nappi-Witten algebra \eqref{invtENW}, while the MENt ones proportional to the $\beta$'s have origin in $\gamma_2$. Such feature is an additional advantage of considering the semigroup expansion method in the NR context. In particular, as the MENt algebra (\ref{MEB})-(\ref{MENt}), the MENt CS gravity action constructed using the $S$-expansion procedure contains also the MEB gravity action \cite{Aviles:2018jzw} as a sub-case.

It is interesting to note that the semigroup chosen is the same used to obtain the MEB algebra from the Nappi-Witten algebra \cite{Concha:2019lhn,Penafiel:2019czp}. Furthermore, as we have shown in Section 3, the same semigroup is used to obtain its
respective relativistic algebra from the $\mathfrak{iso}\left( 2,1\right) \,
\oplus \, \mathfrak{u}\left( 1\right) $ algebra. The fact that the same semigroup can be used in two diverse regimes has already appeared in other contexts. Indeed, the same semigroup can be used to relate diverse finite- and infinite-dimensional Lie (super)algebras \cite{Caroca:2017onr,Caroca:2018obf,Caroca:2019dds} or higher-spin Lie algebras \cite{Caroca:2017izc}.

On the other hand, considering a different semigroup allows us to find a
different NR algebra. One can show that the EEN algebra can alternatively be obtained by applying a $S_{%
\mathcal{M}}^{\left( 2\right) }$-expansion of the enhanced Nappi-Witten
algebra. Indeed, considering the multiplication law of the $S_{\mathcal{M}%
}^{\left( 2\right) }=\left\{ \lambda _{0},\lambda _{1},\lambda _{2}\right\} $
semigroup given by
\begin{equation}
\lambda _{\alpha }\lambda _{\beta }=\left\{
\begin{array}{lcl}
\lambda _{\alpha +\beta }\,\,\,\, & \mathrm{if}\,\,\,\,\alpha +\beta \leq
2 \, ,  &  \\
\lambda _{\alpha +\beta -2}\,\,\, & \mathrm{if}\,\,\,\,\alpha +\beta >2\,, &
\end{array}%
\right.
\end{equation}%
one obtains an $S$-expanded algebra whose generators are related to the
enhanced Nappi-Witten ones as in (\ref{def}) and satisfy the EEN algebra (\ref{MEB}), (\ref{MENt}), (\ref{EEB}), and (\ref{EEN}). Furthermore, considering the Theorem VII. of \cite{Izaurieta:2006zz}, one can show that
the non-vanishing components of the invariant tensor for the EEN algebra are
given not only by (\ref{EENinvt}) but also by the EEB ones (\ref{EEBinvt}). As discussed in previous sections, both of the invariant tensors cannot be obtained from a unique NR limit. Here, the $S$-expansion gives us the complete set of invariant tensors considering only one semigroup, and allowing to write an EEN CS gravity action containing the EEB gravity \cite{Concha:2019lhn} as a sub-case. In particular, the arbitrary constants $\alpha $'s and $\beta $'s are
related to the original ones as in (\ref{alpha}) and (\ref{beta}),
respectively.

As an ending remark, let us mention that one could obtain new NR algebras by
expanding the enhanced Nappi-Witten algebra using the same semigroup used to
obtain their respective relativistic algebras from the $\mathfrak{iso}\left(
2,1\right) \, \oplus \, \mathfrak{u}\left( 1\right) $ algebra. It would be
interesting to extend the present results to supersymmetric extensions of
NR gravity theories.

\section{Discussion}

In this work we have presented a Maxwellian version of the three-dimensional extended Newtonian gravity theory introduced in \cite{Ozdemir:2019orp}. We have shown that the MENt algebra appears as an IW contraction of the [enhanced Maxwell] $\oplus\,\mathfrak{u}(1)\oplus\mathfrak{u}(1)\oplus\mathfrak{u}(1)$ algebra. The additional $U(1)$ gauge fields are necessary to have a finite and non-degenerate invariant tensor allowing the proper construction of a NR CS action. Furthermore, we have also applied the NR limit at the level of the CS action.  We have then extended our results to accommodate a cosmological constant in the MENt gravity theory. To this end we have presented a new NR symmetry which can be seen as an enlargement of the extended Newtonian algebra and we have denoted it as EEN algebra. Such novel NR symmetry can be rewritten either as three copies of the enhanced Nappi-Witten algebra defined in \cite{Bergshoeff:2020fiz} or as the direct sum of the exotic Newtonian algebra introduced in \cite{Concha:2019dqs} with the enhanced Nappi-Witten one. 

We have also explored an alternative procedure to recover our results by considering the semigroup expansion of the enhanced Nappi-Witten symmetry. In particular, we have shown that the same semigroup used to relate relativistic algebras can be adopted to reproduce their NR counterparts. In addition, we have pointed out that the vanishing cosmological constant limit appearing at the relativistic level is inherited in the NR version.

\begin{figure}[htp]
    \centering
    \includegraphics[width=14cm]{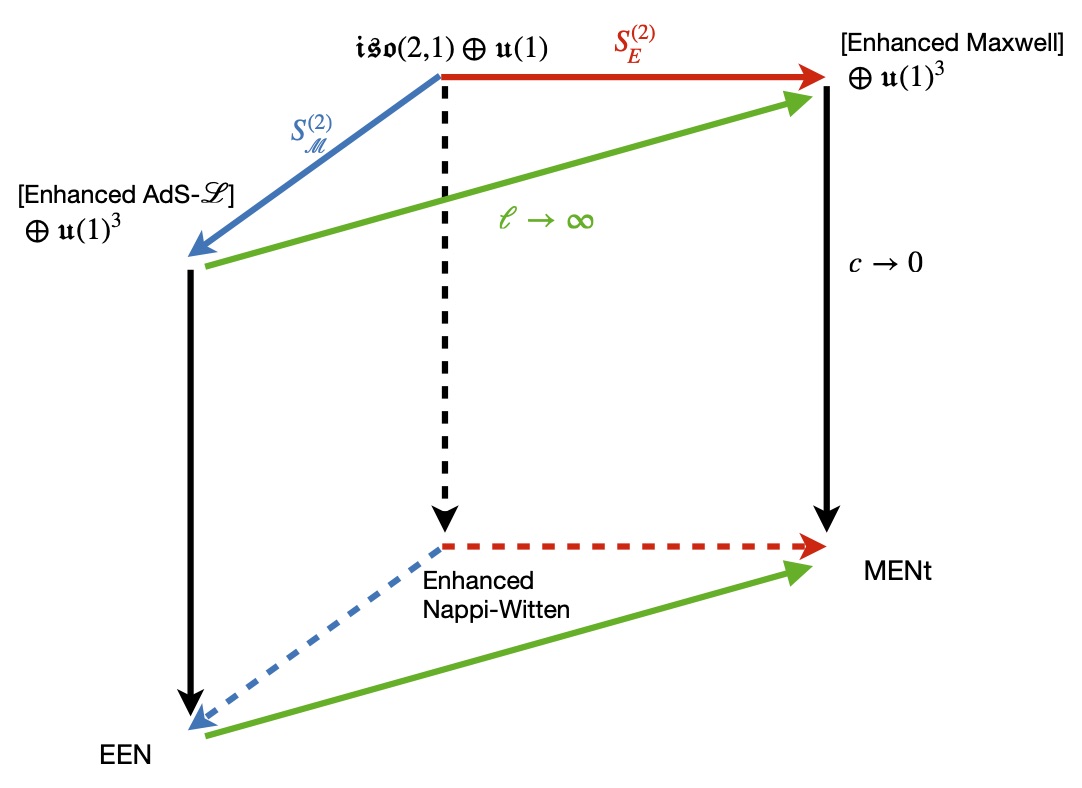}
    \caption{Expansions and limits}
    \label{fig:Sexp}
\end{figure}

The $S$-expansion method has the advantage to provide not only the new Newtonian algebras but also to reproduce the most general invariant tensor for the respective NR algebras. Interestingly, the MENt and EEN gravity theories constructed with expanded invariant tensor contain known extended Bargmann gravity actions as sub-cases. One can notice that the scenario graphically displayed in Figure \ref{fig:Sexp} also holds for the particular sub-cases. In particular, as was noticed in \cite{Concha:2019lhn}, the MEB and the EEB algebras can alternatively be obtained as $S$-expansions of the Nappi-Witten algebra.

Let us stress that our novel CS models can be seen as generalizations of the extended Newtonian gravity \cite{Hansen:2018ofj}, and thus they differ from the standard Newtonian gravity for which no action is known. Although we did not approach in the present work the matter coupling, we could argue that as in the extended Bargmann gravity and extended Newtonian gravity cases, the equations of motion derived form our models coupled with matter would allow a wider class of background geometries than the ones presented in standard Newtonian gravity. Besides, also in our case matter would source all components of the Riemann tensor. Consequently, the MENt and EEN gravities would admit backgrounds with non-trivial curvature whenever matter is present similarly as it happens in the case of the extended Bargmann gravity and extended Newtonian gravity. Then, as a consequence of our Maxwellian generalization, we would obtain a modified version of the Poisson equation as equation of motion.

Our results could be seen as a starting point for diverse studies. It would be interesting to explore the extensions of our analysis to supergravity. Although the large amount of NR gravity models, the supersymmetric extension of NR gravity theories has only been approached recently \cite{Andringa:2013mma,Bergshoeff:2015ija,Bergshoeff:2016lwr,Ozdemir:2019orp,Ozdemir:2019tby,Concha:2019mxx} due to the difficulty to define a proper NR limit. One way to avoid such problem could be through the $S$-expansion method considered here. Indeed, one could conjecture that Figure \ref{fig:Sexp} can be generalized to the presence of supersymmetry. A supersymmetric action for the Maxwellian and enlarged version of the extended Newtonian superalgebra could serve to approach the construction of NR field theories on curved background by means of localization \cite{Festuccia:2011ws,Pestun:2007rz,Marino:2012zq}. The supersymmetric extension of the Maxwellian extended Newtonian gravity and its exotic version shall be approached in a future work. 

At the relativistic level, the Maxwell symmetries describe a relativistic particle in a constant electromagnetic field background \cite{Schrader:1972zd,Bacry:1970ye}. Here, we have shown that an enhancement of the Maxwell algebra allows us to establish a well-defined IW contraction producing the Maxwellian version of the extended Newtonian algebra. It would be interesting to analyze the physical interpretation by studying the consequences of the additional gauge fields and central charges of the MENt CS gravity theory and by analyzing the geodesic equation for a massive particle from a gauge theory point of view. On the other hand, it would be compelling to explore the coupling of well-known relativistic matter systems to the Newtonian gravity theories presented here. For example, one could consider a relativistic point particle.

It would be worth it to study the generalization of the results presented here to the ultra-relativistic regime. In particular, a Maxwellian version of the three-dimensional Carroll gravity theory \cite{Bergshoeff:2016soe,Bergshoeff:2017btm} could be constructed analogously to the MEB one. Then, a cosmological constant could be accommodated in a very similar way to the one considered in the AdS-Caroll CS gravity \cite{Matulich:2019cdo,Ravera:2019ize,Ali:2019jjp}. Another natural extension of our results is the generalization of the obtained NR symmetries to four and higher spacetime dimensions.

Another point that deserves further investigation is the physical implications of the additional gauge fields appearing in the enhanced Maxwell and AdS-$\mathcal{L}$ gravity theories introduced here. As it was discussed in \cite{Concha:2018zeb}, the additional gauge generator $Z_{A}$ appearing in the Maxwell algebra influences not only the asymptotic sector but also its vacuum energy and vacuum angular momentum. On the other hand, the boundary dynamics of the AdS-$\mathcal{L}$ CS gravity is described by three copies of the Virasoro algebra \cite{Concha:2018jjj}. One may then ask how the conserved charges and solutions are modified by the presence of the additional set of relativistic generators $\{S_{A},T_{A},V_{A} \}$.

\section*{Acknowledgments}

This work was funded by the National Agency for Research and Development ANID (ex-CONICYT) - PAI grant No. 77190078 (P.C.) and the FONDECYT Project N$^{\circ }$3170438 (E.R.). P.C. would like to
thank to the Dirección de Investigación and Vice-rectoría de Investigación
of the Universidad Católica de la Santísima Concepción, Chile, for their
constant support. L.R. would like to thank the Department of Applied Science and Technology of the Polytechnic University of Turin, and in particular Professors F. Dolcini and A. Gamba, for financial support.

\appendix

\section{Curvature two-forms}
The curvature two-forms for the EEN algebra are given by
\begin{eqnarray}
R\left( \omega \right) &=&d\omega \,,  \notag \\
R^{a}\left( \omega ^{b}\right) &=&d\omega ^{a}+\epsilon ^{ac}\omega
\omega_{c}\,,  \notag \\
R\left( \tau \right) &=& d\tau \,,  \notag \\
\hat{R}^{a}\left( e^{b}\right) &=& d e^a + \epsilon ^{ac}\omega e_{c} + \epsilon
^{ac}\tau \omega_{c} + \frac{1}{\ell^2} \epsilon^{ac} \tau k_c + \frac{1}{\ell^2} \epsilon^{ac} k e_c \,,  \notag \\
R\left( k\right) &=& dk \,, \notag \\
\hat{R}^{a}\left( k^{b}\right) &=& d k^a + \epsilon ^{ac}\omega k_{c} + \epsilon
^{ac}\tau e_{c} + \epsilon ^{ac} k \omega_{c} + \frac{1}{\ell^2} \epsilon^{ac} k k_c \,,  \notag \\
\hat{R}\left( m\right) &=& dm + \epsilon ^{ac} e_a \omega_{c} + \frac{1}{\ell^2} \epsilon^{ac} e_a k_c \,,  \notag \\
R\left( s\right) &=&ds+\frac{1}{2}\epsilon ^{ac}\omega _{a}\omega _{c}\,, \notag \\
\hat{R}\left( t\right) &=& dt + \epsilon ^{ac}\omega_{a} k_{c} + \frac{1}{2}
\epsilon ^{ac} e_{a} e_{c} + \frac{1}{2\ell^2} \epsilon^{ac} k_a k_c \,, \notag \\
\hat{R}^a\left( t^b\right) &=& dt^a + \epsilon ^{ac}\omega t_{c} + \epsilon^{ac} \tau b_c + \epsilon^{ac} s e_c + \epsilon^{ac} m \omega_c + \frac{1}{\ell^2} \epsilon^{ac} \tau v_c + \frac{1}{\ell^2} \epsilon^{ac} k t_c \notag \\
&& + \frac{1}{\ell^2} \epsilon^{ac} m k_c + \frac{1}{\ell^2} \epsilon^{ac} t e_c \,, \notag \\
R^a\left( b^b\right) &=& db^a + \epsilon ^{ac}\omega b_{c} + \epsilon^{ac} s \omega_c \,, \notag \\
\hat{R}^a\left( v^b\right) &=& dv^a + \epsilon ^{ac}\omega v_{c} + \epsilon^{ac} \tau t_c + \epsilon^{ac} k b_c + \epsilon^{ac} s k_c + \epsilon^{ac} m e_c + \epsilon^{ac} t \omega_c \notag \\
&& + \frac{1}{\ell^2} \epsilon^{ac} k v_c + \frac{1}{\ell^2} \epsilon^{ac} t k_c \,, \notag \\
R\left( b\right) &=& db + \epsilon ^{ac} \omega_a b_c \,,  \notag \\
\hat{R}\left( y\right) &=& dy + \epsilon ^{ac} \omega_a t_c + \epsilon^{ac} b_a e_c + \frac{1}{\ell^2} \epsilon^{ac} e_a v_c + \frac{1}{\ell^2} \epsilon^{ac} k_a t_c \,,  \notag \\
\hat{R}\left( \varpi\right) &=& d\varpi + \epsilon ^{ac} \omega_a v_c + \epsilon^{ac} t_a e_c + \epsilon^{ac} k_a b_c + \frac{1}{\ell^2} \epsilon^{ac} k_a v_c \,.  \label{curvEEN}
\end{eqnarray}
The vanishing cosmological constant limit $\ell \rightarrow \infty $
naturally reproduces the curvature two-forms of the MENt algebra. In particular, in the flat limit, the curvature two-forms with an hat reduce to the MENt ones \eqref{curvMENt}.

\section{Gauge transformations}
The gauge transformations for the EEN gauge fields are given by
\begin{eqnarray}
\delta \omega &=&d\Omega \,,  \notag \\
\delta \omega^a &=&d\Omega ^{a}+\epsilon ^{ac}\omega \Omega_{c} - \epsilon ^{ac}\Omega \omega_{c}\,,  \notag \\
\delta \tau &=& d \Lambda \,,  \notag \\
\delta e^a &=& d \Lambda^a + \epsilon ^{ac}\omega \Lambda_{c} - \epsilon ^{ac}\Omega e_{c} + \epsilon ^{ac}\tau \Omega_{c} - \epsilon ^{ac}\Lambda \omega_{c} + \frac{1}{\ell^2} \epsilon^{ac} \tau \kappa_c - \frac{1}{\ell^2} \epsilon^{ac} \Lambda k_c \notag \\
&& + \frac{1}{\ell^2} \epsilon^{ac} k \Lambda_c - \frac{1}{\ell^2} \epsilon^{ac} \kappa e_c \,,  \notag \\
\delta k &=& d \kappa \,, \notag \\
\delta k^a &=& d \kappa^a + \epsilon ^{ac}\omega \kappa_{c} - \epsilon ^{ac}\Omega k_{c} + \epsilon^{ac}\tau \Lambda_{c} - \epsilon^{ac}\Lambda e_{c} + \epsilon ^{ac} k \Omega_{c} - \epsilon ^{ac} \kappa \omega_{c} + \frac{1}{\ell^2} k \kappa_c - \frac{1}{\ell^2} \kappa k_c  \,,  \notag \\
\delta m &=& d \chi + \epsilon ^{ac} e_a \Omega_{c} - \epsilon ^{ac} \Lambda_a \omega_{c} + \frac{1}{\ell^2} \epsilon^{ac} e_a \kappa_c - \frac{1}{\ell^2} \epsilon^{ac} \Lambda_c k_c \,,  \notag \\
\delta s &=& d \varsigma + \epsilon ^{ac}\omega _{a}\Omega _{c}\,, \notag \\
\delta t &=& d \pi + \epsilon ^{ac}\omega_{a} \kappa_{c} - \epsilon ^{ac}\Omega_{a} k_{c} + \epsilon ^{ac} e_{a} \Lambda_{c} + \frac{1}{\ell^2} \epsilon^{ac} k_a \kappa_c \,, \notag \\
\delta t^a &=& d \pi^a + \epsilon ^{ac}\omega \pi_{c} - \epsilon ^{ac}\Omega t_{c} + \epsilon^{ac} \tau \rho_c - \epsilon^{ac} \Lambda b_c + \epsilon^{ac} s \Lambda_c - \epsilon^{ac} \varsigma e_c + \epsilon^{ac} m \Omega_c - \epsilon^{ac} \chi \omega_c \notag \\
&& + \frac{1}{\ell^2} \epsilon^{ac} \tau \nu_c - \frac{1}{\ell^2} \epsilon^{ac} \Lambda v_c + \frac{1}{\ell^2} \epsilon^{ac} k \pi_c - \frac{1}{\ell^2} \kappa t_c + \frac{1}{\ell^2} \epsilon^{ac} m \kappa_c - \frac{1}{\ell^2} \epsilon^{ac} \chi k_c + \frac{1}{\ell^2} \epsilon^{ac} t \Lambda_c - \frac{1}{\ell^2} \pi e_c \,, \notag \\
\delta b^a &=& d \rho^a + \epsilon ^{ac}\omega \rho_{c} - \epsilon ^{ac}\Omega b_{c} + \epsilon^{ac} s \Omega_c - \epsilon^{ac} \varsigma \omega_c \,, \notag \\
\delta v^a &=& d \nu^a + \epsilon ^{ac}\omega \nu_{c} - \epsilon ^{ac}\Omega v_{c}  + \epsilon^{ac} \tau \pi_c - \epsilon^{ac} \Lambda t_c + \epsilon^{ac} k \rho_c - \epsilon^{ac} \kappa b_c + \epsilon^{ac} s \kappa_c - \epsilon^{ac} \varsigma k_c \notag \\
&& + \epsilon^{ac} m \Lambda_c - \epsilon^{ac} \chi e_c + \epsilon^{ac} t \Omega_c - \epsilon^{ac} \pi \omega_c + \frac{1}{\ell^2} \epsilon^{ac} k \nu_c - \frac{1}{\ell^2} \epsilon^{ac} \kappa v_c + \frac{1}{\ell^2} \epsilon^{ac} t \kappa_c - \frac{1}{\ell^2} \epsilon^{ac} \pi k_c \,, \notag \\
\delta b &=& d \rho + \epsilon ^{ac} \omega_a \rho_c - \epsilon ^{ac} \Omega_a b_c \,,  \notag \\
\delta y &=& d \gamma + \epsilon ^{ac} \omega_a \pi_c - \epsilon ^{ac} \Omega_a t_c + \epsilon^{ac} b_a \Lambda_c - \epsilon^{ac} \rho_a e_c + \frac{1}{\ell^2} \epsilon^{ac} e_a \nu_c - \frac{1}{\ell^2} \epsilon^{ac} \Lambda_a v_c \notag \\
&& + \frac{1}{\ell^2} \epsilon^{ac} k_a \pi_c - \frac{1}{\ell^2} \epsilon^{ac} \kappa_a t_c \,,  \notag \\
\delta \varpi &=& d \varrho + \epsilon ^{ac} \omega_a \nu_c - \epsilon ^{ac} \Omega_a v_c + \epsilon^{ac} t_a \Lambda_c - \epsilon^{ac} \pi_a e_c + \epsilon^{ac} k_a \rho_c - \epsilon^{ac} \kappa_a b_c \notag \\
&& + \frac{1}{\ell^2} \epsilon^{ac} k_a \nu_c - \frac{1}{\ell^2} \epsilon^{ac} \kappa_a v_c \,. \label{gaugetrEEN}
\end{eqnarray}
Let us note that, in the limit $\ell \rightarrow \infty$, \eqref{gaugetrEEN} reproduces the gauge transformations for the MENt gauge fields \eqref{connMENt}.

\end{document}